\documentclass[11pt]{article}
	
	\newcommand{\blind}{0}
	
    \makeatletter
    \renewcommand\section{\@startsection {section}{1}{\z@}%
                                       {-3.5ex \@plus -1ex \@minus -.2ex}%
                                       {2.3ex \@plus.2ex}%
                                       {\normalfont\fontfamily{phv}\fontsize{16}{19}\bfseries}}
    \renewcommand\subsection{\@startsection{subsection}{2}{\z@}%
                                         {-3.25ex\@plus -1ex \@minus -.2ex}%
                                         {1.5ex \@plus .2ex}%
                                         {\normalfont\fontfamily{phv}\fontsize{14}{17}\bfseries}}
    \renewcommand\subsubsection{\@startsection{subsubsection}{3}{\z@}%
                                        {-3.25ex\@plus -1ex \@minus -.2ex}%
                                         {1.5ex \@plus .2ex}%
                                         {\normalfont\normalsize\fontfamily{phv}\fontsize{14}{17}\selectfont}}
    \makeatother
	
	\usepackage{amsmath}
	\usepackage{graphicx}
	\usepackage{enumerate}
	\usepackage{xcolor}
    \usepackage{natbib}
	\usepackage{url}
	
        \usepackage[affil-it]{authblk}
        \usepackage{amssymb}
        \usepackage{amsthm}
        \usepackage{bm}
        \usepackage{graphicx}
        \usepackage{subcaption}
        \usepackage{graphicx}
        \usepackage{booktabs}
        \newtheorem{theorem}{Theorem}[section]

\begin{document}
		
		\def\spacingset#1{\renewcommand{\baselinestretch}%
			{#1}\small\normalsize} \spacingset{1}
		
		\if0\blind
		{
            \title{\bf 3D-SONAR: Self-Organizing Network for 3D Anomaly Ranking}
            \author[1]{Guodong Xu}
            \author[1,2,3]{Juan Du\thanks{Corresponding author. E-mail: juandu@hkust-gz.edu.cn}}
            \author[4]{Hui Yang}
            \affil[1]{Smart Manufacturing Thrust, The Hong Kong University of Science and Technology (Guangzhou), Guangzhou, China}
            \affil[2]{Department of Mechanical and Aerospace Engineering, The Hong Kong University of Science and Technology, Hong Kong SAR, China}
            \affil[3]{Academy of Interdisciplinary Studies, The Hong Kong University of Science and Technology, Hong Kong SAR, China}
            \affil[4]{Complex Systems Monitoring, Modeling and Control Laboratory, The Pennsylvania State University, University Park, PA, USA}
            \date{}
			\maketitle
		} \fi
		
		\if1\blind
		{
            \title{\bf 3D-SONAR: Self-Organizing Network for 3D Anomaly Ranking}
			\author{Author information is purposely removed for double-blind review}
            \date{}
            \maketitle
			\bigskip
		} \fi
		\bigskip
		
	\begin{abstract}
    Surface anomaly detection using 3D point cloud data has gained increasing attention in industrial inspection. However, most existing methods rely on deep learning techniques that are highly dependent on large-scale datasets for training, which are difficult and expensive to acquire in real-world applications. To address this challenge, we propose a novel method based on self-organizing network for 3D anomaly ranking, also named 3D-SONAR. The core idea is to model the 3D point cloud as a dynamic system, where the points are represented as an undirected graph and interact via attractive and repulsive forces. The energy distribution induced by these forces can reveal surface anomalies. Experimental results show that our method achieves superior anomaly detection performance in both open surface and closed surface without training. This work provides a new perspective on unsupervised inspection and highlights the potential of physics-inspired models in industrial anomaly detection tasks with limited data.
	\end{abstract}
			
	\noindent%
	{\it Keywords:}  Anomaly detection; 3D point cloud; Self-organizing network; Untrained method.

	\spacingset{1.5} 

\section{Introduction} \label{s:intro}

In the field of manufacturing, surface quality is critical to determining product performance, reliability, and appearance. Surface anomalies such as scratches, dents, and holes not only affect aesthetics but may also compromise structural integrity, leading to functional failures. Surface anomaly detection is essential for ensuring surface quality and are widely applied in various industrial scenarios ~\citep{liIndustrialapplicationoriented2025,duPosition2025,raniAdvancements2024}. Recent advances in 3D sensing technology~\citep{haleemExploring2022} have enabled the acquisition of high-resolution point cloud data, providing new opportunities for surface inspection. Compared to 2D images, 3D scanners can provide precise spatial coordinates, enabling quantitative 3D geometric characterization about anomalies such as maximum depth and orientation.

In personalized and flexible production lines, anomaly data is often scarce, and there is a growing demand for single-sample anomaly detection methods. At the same time, the surfaces of many man-made products can be assumed to be piecewise smooth, such as airplane wings and historical artifacts. Our aim is to develop an untrained anomaly detection method for products with piecewise-smooth surfaces, with a focus on identifying anomalies such as scratches, dents, and holes. Untrained method~\citep{taoAnomalyDetectionFabricated2023,taoPointSGRADESparseLearning2025,Ye06082025} refers to one that can directly detect anomalies in a single sample without relying on any training dataset, including both anomalous and anomaly-free data.

Recently, untrained anomaly detection attracts more attentions and can be broadly categorized into two main paradigms. The first paradigm relies on local geometric descriptors (such as surface normal vectors or curvatures) to identify anomalies using handcrafted features. Representative methods include Fast Point Feature Histograms (FPFH)~\citep{rusuFast2009,miaoPipeline2022} and curvature-based analysis method~\citep{jovancevic3D2017}. While effective in certain environments, these approaches perform poorly in regions with high curvatures and non-smooth geometry.
The second paradigm adopts a global modeling strategy that leverages strong geometric priors to reconstruct a reference surface, thereby enabling anomaly identification. For instance,~\cite{taoAnomalyDetectionFabricated2023} proposed a probabilistic Bayesian network framework for anomaly detection under the assumption that the point cloud is globally smooth. Subsequently,~\cite{taoPointSGRADESparseLearning2025} formulated the problem as a sparse learning task to capture localized sparse anomalies. This work has been further extended to periodic image analysis through self-representation models~\citep{Ye06082025}. However, these approaches are only applicable when the target object is either globally smooth or exhibits a periodic structure. 

Consequently, there are several challenges in the design and development of untrained methods for anomaly detection on complex surfaces, as follows:

\begin{enumerate}
  \item Complex surfaces often contain a range of diverse anomalies, which requires highly generalizable detection criteria for effective identification.
  \item The proposed method is training-free and performs direct inference from a single sample, without requiring large-scale normal datasets or annotated data.
  \item To be applicable to high-density point cloud data, the proposed methodology should be computationally efficient.
\end{enumerate}

To address these challenges, we propose an untrained surface anomaly detection method for point cloud data based on the self-organizing network (SON)~\citep{yangSelforganizing2020}. SON was originally proposed to uncover the underlying geometric structure of complex networks through physics-based simulation. By analyzing the energy distribution established by SON, we observe that high energy regions often correspond to anomalous areas. Meanwhile, we theoretically derive the relationship between the energy established by SON and surface anomalies, and present the corresponding theorem in the main text. Based on these observations and supported by mathematical analysis, we leverage SON to develop an untrained anomaly detection method named SONAR (Self-Organizing Network for 3D
anomaly ranking). Our main contributions are as follows:

\begin{enumerate}
    \item We propose an untrained anomaly detection method based on SON, capable of performing detection directly on a single piecewise-smooth surface sample without additional anomaly-free data or annotated data.
    \item Our approach avoids normal vector estimation, significantly improving anomaly detection performance in sharp and high-curvatures regions.
    \item We introduce a geometric local energy normalization scheme and energy boundary conditions to address a common challenge, i.e., boundaries and sharp edges are often misidentified as anomalies.
\end{enumerate}

The remainder of this paper is organized as follows: 
Section~\ref{s:LR} reviews the literature on anomaly detection based on 3D point cloud data. 
Section~\ref{s:methods} presents the construction of the proposed SONAR for point cloud surface anomaly detection.
Section~\ref{s:case study} evaluates the performance of the proposed method using both synthetic and real-world datasets.
Finally, this paper is concluded in Section~\ref{s:conclusion}.

\section{Literature review} \label{s:LR}
Based on whether anomaly detection methods require anomaly-free training data, we divide existing unsupervised anomaly detection method into two categories: training-based unsupervised methods and untrained methods. As for untrained methods, we further categorize them based on the scale of geometric information, dividing them into local and global geometric feature-based methods.

\subsection{\emph{Training methods based on anomaly-free samples}}

Unsupervised anomaly detection methods are trained on anomaly-free samples. Anomaly detection in such frameworks can be primarily achieved through two types of techniques, namely feature embedding and reconstruction.

Feature embedding methods generally consist of two stages: (1) extracting latent features from anomaly-free training data, and (2) modeling the distribution of normal features using techniques such as Memory Bank~\citep{du3DVisionbasedAnomaly2025} or Knowledge distillation (KD)~\citep{gouKnowledge2021}, which are then used to detect surface anomalies during inference. Examples include:~\cite{zhaoPointCore2024} proposed PointCore, an unsupervised point cloud anomaly detection framework inspired by PatchCore~\citep{rothTotal2022}, where both local (coordinates) and global (PointMAE~\citep{pangMasked2022}) representations are stored in a memory bank to capture the distribution of normal features. M3DM~\citep{wangMultimodal2023} utilizes a pre-trained encoder, PointMAE~\citep{pangMasked2022}, to extract latent features from 3D point cloud data. \cite{bergmannAnomaly2023} proposed a self-supervised teacher-student framework, where the teacher is built through local receptive field reconstruction and learns expressive representations. The student then learns to emulate the teacher on anomaly-free data, enabling precise localization of anomalies during inference based on regression errors. A key challenge in this approach is the student network's tendency to over-generalize, potentially leading to inaccurate reconstruction. To address this issue,~\cite{rudolphAsymmetric2023a} introduced an asymmetric architecture, using a normalizing flow as the teacher and a traditional Convolutional Neural Networks(CNN) as the student, effectively mitigating overfitting.

Unlike feature-based approaches, reconstruction methods perform anomaly detection by training an autoencoder to reconstruct the input 3D point cloud based on learned data patterns. The difference between the input and the reconstructed output is then used as an anomaly score to identify anomalous features. Numerous approaches are built upon AutoEncoder (AE) architectures~\citep{liComprehensive2023}. For instance, EasyNet~\citep{chenEasyNet2023} employs a multi-scale, multi-modality feature encoder-decoder for 3D depth map reconstruction. Cheating Depth~\citep{zavrtanikCheating2024} proposed 3DSR, a 3D surface anomaly detection method combining RGB and depth information through a depth-aware discrete autoencoder, along with a simulation-based depth data augmentation strategy to enhance 3D surface anomaly detection. However, these methods are often limited to structured point clouds. To address this limitation,~\cite{liScalable2024} proposed IMRNet, a self-supervised iterative mask-based reconstruction network for reconstructing unstructured point clouds and improving the accuracy of anomaly detection.

Recently, diffusion-based~\citep{hoDenoising2020} reconstruction techniques have also been applied to 3D anomaly detection. These methods leverage the powerful generative capabilities of diffusion models to reconstruct the input point cloud, and then compare the reconstructed result with the original input to enable anomaly detection. For instance, R3D-AD~\citep{zhouR3DAD2024} leverages a conditional diffusion model to reconstruct anomaly-free point clouds and detects anomalies by comparing the input and output. RDDPM~\citep{moradiRDDPM2025} introduces the Huber loss~\citep{meyerAlternative2020} into diffusion model, enabling unsupervised anomaly detection even in the presence of contaminated data.

However, these methods require a large amount of anomaly-free data. In personalized and flexible production lines, such data may be scarce because each product variant is produced in low volumes or even as a one-off, limiting the availability of representative samples.

\subsection{Untrained methods}
Unlike the two types of methods mentioned above, untrained methods are capable of processing single samples directly, without requiring any anomaly-free training data.

\subsubsection{\emph{Local geometric methods}}
Currently, the majority of untrained methods rely on local geometric information for anomaly detection. For example,~\cite{jovancevic3D2017} developed a region-growing segmentation approach based on surface normal vectors and curvatures information to identify anomalies on airplane exterior surfaces.~\cite{hitchcoxRandom2018} proposed a graph-based segmentation method to label anomalous points, although it requires manual selection of initial anomaly seeds.~\cite{miaoPipeline2022} employed the FPFH~\citep{rusuFast2009}, a 3D descriptor to identify dentify anomalous points. However, these local geometry-based methods are heavily dependent on normal vector estimation, which performs poorly in regions with high curvatures and non-smooth geometry.

\subsubsection{\emph{Global geometric Methods}}
These methods detect anomalies by explicitly reconstructing a reference surface. \cite{taoAnomalyDetectionFabricated2023} proposed a method that models the reference surface as a B-spline surface over a parametric base plane. However, this approach is limited in representing more general free-form surfaces, such as closed surfaces. Subsequently,~\cite{taoPointSGRADESparseLearning2025} decoupled the point cloud into three components: the reference plane (the original, anomaly-free surface), anomalous points, and noise points. They proposed a sparse learning framework and formulated the anomaly detection task as an optimization problem with penalty terms. However, these methods are only applicable to surface anomaly detection under the assumption of smooth surface.

In addition, some untrained approaches exploit strong priors such as rotational symmetry~\citep{10.1115/1.4068472} or periodicity~\citep{Ye06082025} to detect anomalies. Although these methods perform well in specific scenarios, they struggle to generalize to more complicated surfaces without such properties.

In summary, the aforementioned methods cannot simultaneously achieve being untrained, normal-vector-estimation-free, and applicable to non-smooth surfaces. Therefore, we propose the SONAR framework to fill the research gaps.

\section{Methodology} \label{s:methods}
In this section, we introduce the proposed SONAR in detail to address the aforementioned challenges and achieve accurate 3D surface anomaly detection without training, focusing on common anomalies such as scratches, dents and holes. Furthermore, we make two assumptions:

\begin{enumerate}
    \item Anomalies are sparse. This assumption is highly consistent with real-world industrial production scenarios.
    
    \item The surface of point cloud is assumed to be piecewise-smooth with limited overall curvatures, a common characteristic in many real-world industrial products.
\end{enumerate}

\subsection{
\emph{Overview of SONAR methodology}} \label{s:methods.1}
Given an input 3D point cloud, we represent it as a matrix $\bm{V} = [\bm{v}_1, \dots, \bm{v}_N]^\top \in \mathbb{R}^{N \times 3}$, where each $\bm{v}_i \in \mathbb{R}^3$ denotes the 3D coordinates of the $i$-th point in the point cloud. Our objective is to detect anomalous surface regions such as scratches, dents, or bumps. Our goal is to obtain the anomaly indicator vector $\bm{a} = [a_1, a_2, \dots, a_N]^T \in \{0,1\}^N$, where each entry $a_i$ corresponds to the $i$-th point $\bm{x}_i$ and indicates its status: $a_i = 1$ signifies that $\bm{x}_i$ is an anomalous point, while $a_i = 0$ indicates that it is normal. Due to the sparsity assumption of anomalies , the vector $\bm{a}$ is typically sparse, with only a small subset of the entries being $1$, and the majority being $0$.

\begin{figure}
    \centering
    \includegraphics[width=\linewidth]{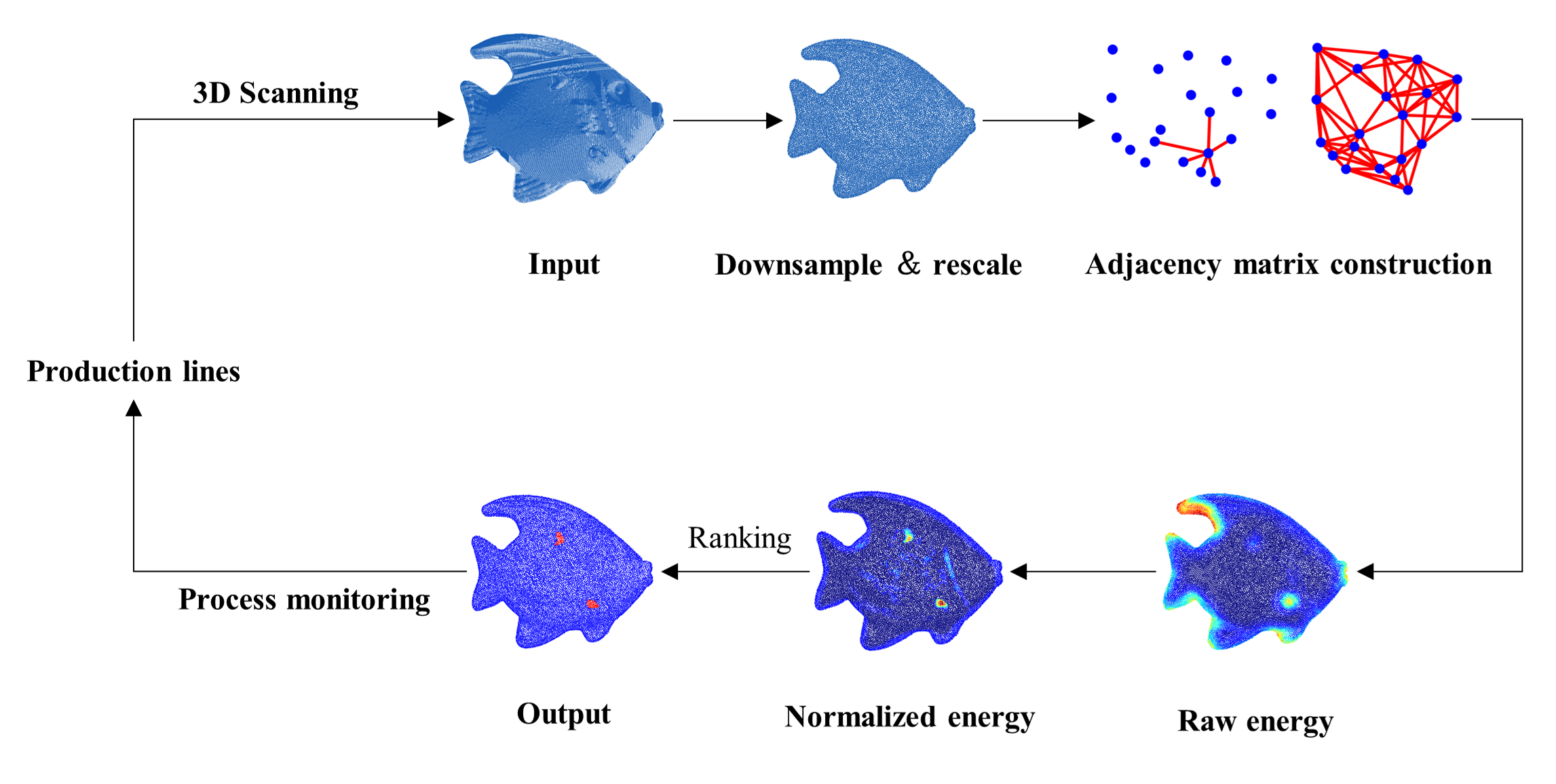}
    \caption{Flowchart of SONAR}
    \label{fig:framework}
\end{figure}

The framework of SONAR is shown in Figure~\ref{fig:framework}. Given point clouds $\bm{V}$ as the input, the final output is an anomaly indicator vector $\bm{a}$. The color transitions from blue (normal) to red (anomalous). To accelerate computation and eliminate the influence of varying input sizes, the point cloud is first downsampled and rescaled into a fixed size. Subsequently, we strictly follow the approach of SON to compute the adjacency matrix and calculate the energy of each point. To focus on anomaly detection, the energy is locally normalized to highlight relatively anomalous regions. Finally, anomaly indicator vector $\bm{a}$ is obtained by ranking the normalized energies.

In practical applications, this pipeline is designed for scalable integration into production lines, where it supports automated process monitoring by flagging potentially defective regions in scanned 3D objects. The resulting heatmaps(output) can be directly consumed by quality control systems or robotic sorting mechanisms, enabling real-time feedback and intervention during manufacturing.

While achieving untrained point cloud anomaly detection, SONAR makes several key contributions: unlike some previous methods that heavily rely on normal vectors, the proposed method does not require normal vector estimation. Additionally, by incorporating an energy boundary condition, it effectively addresses false positive detections at boundaries.

\begin{figure}
    \centering
    \includegraphics[width=\linewidth]{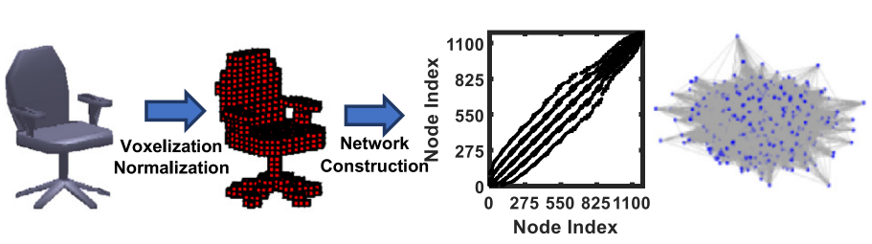}
    \caption{Flowchart of SON~\citep{yangSelforganizing2020}}
    \label{fig:son_network_construction}
\end{figure}

\subsection{\emph{Anomaly detection based on SON}} \label{s:methods.2}
\begin{figure}
\centering
\includegraphics[width=0.8\textwidth]{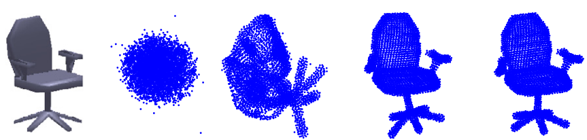}
\caption{Illustration of the self-organizing process~\citep{yangSelforganizing2020}}
\label{fig:SON_self_organizing}
\end{figure}

\begin{figure}
    \centering
    \includegraphics[width=\textwidth]{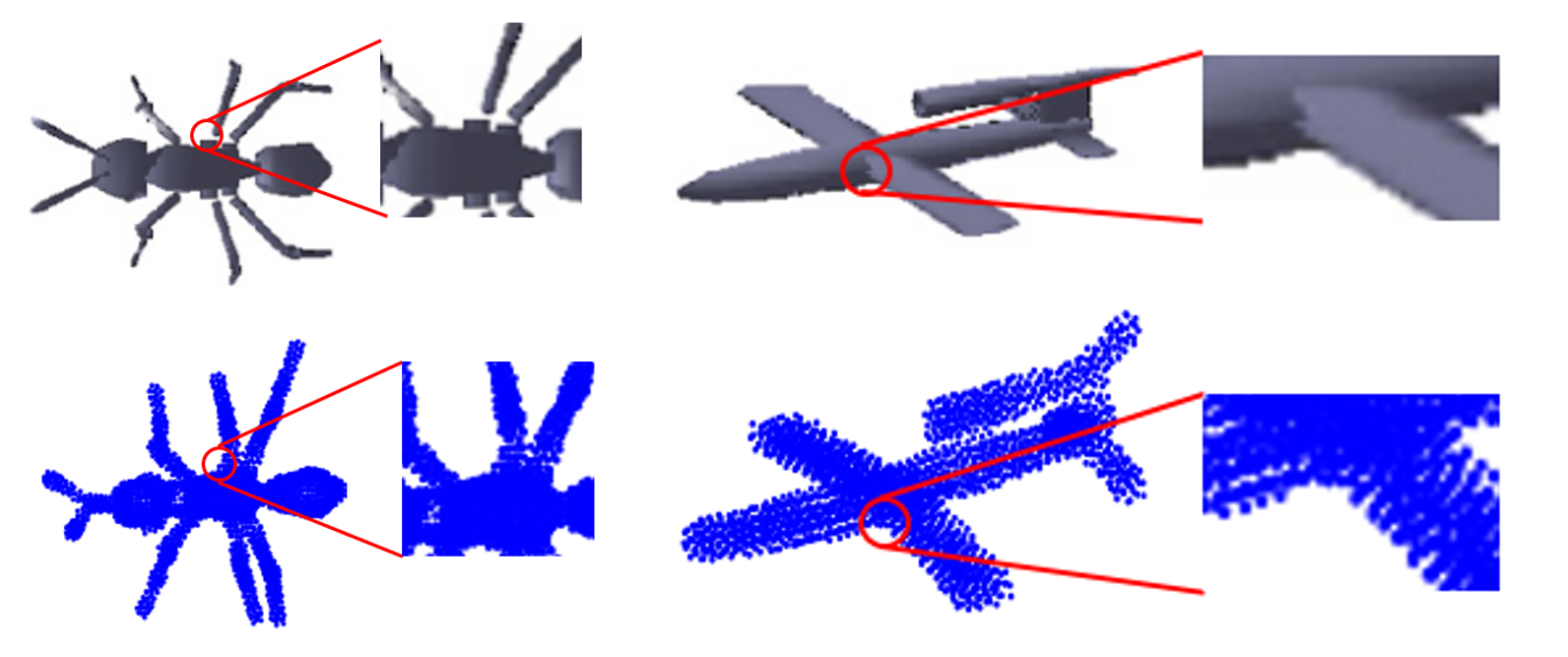}
    \caption{Comparison between the original and reconstructed structures~\citep{yangSelforganizing2020}}
    \label{fig:different_area}
\end{figure}

SON was originally proposed to reveal the intrinsic geometric structure of a network by modeling it as a physical system of springs and charged particles. As shown in Figure.~\ref{fig:son_network_construction}, a chair-shaped point cloud is shuffled after the adjacency matrix has been constructed. As shown in Figure~\ref{fig:SON_self_organizing}, SON can then reconstruct a new point cloud that recovers the original overall chair shape by a force-based optimization method. However, while the final reconstruction accurately recovers most regions, obvious differences appear in certain localized areas—highlighted by red circles in Figure~\ref{fig:different_area}. These differences indicate that the force model constructed by SON becomes unbalanced in regions characterized by high curvatures, discontinuities, or non-smooth geometry. This observation motivates us to further develop an anomaly detection method based on SON.

\subsubsection{\emph{SON preliminaries}} \label{s:methods.2.1}

\begin{figure}
    \centering
    \begin{subfigure}{0.48\textwidth}
        \centering
        \includegraphics[width=\textwidth]{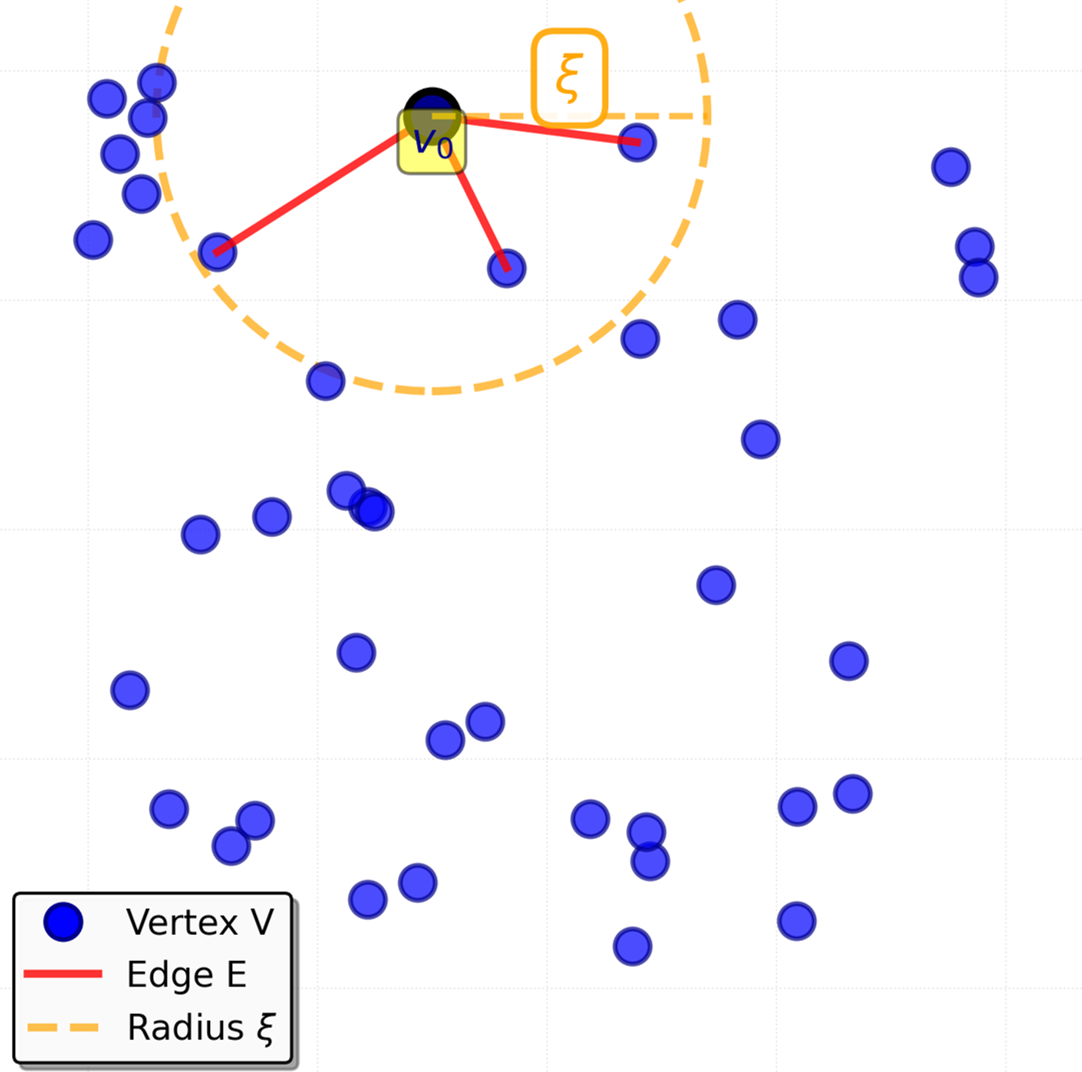}
        \caption{Adjacency diagram for vertex 0}
        \label{fig:step1}
    \end{subfigure}
    \hfill
    \begin{subfigure}{0.48\textwidth}
        \centering
        \includegraphics[width=\textwidth]{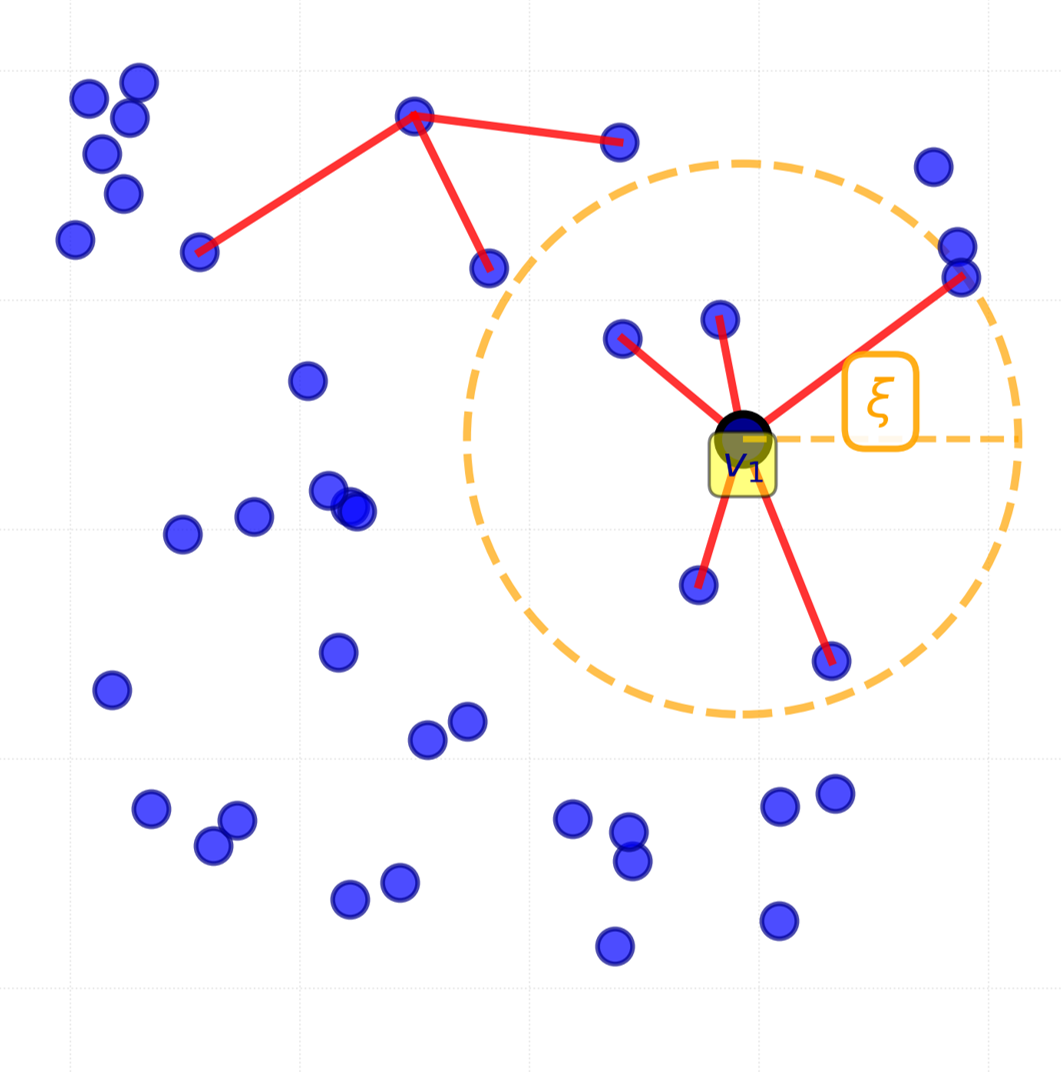}
        \caption{Adjacency diagram for vertex 1}
        \label{fig:step2}
    \end{subfigure}
    \caption{Illustration of the recurrence network construction.}
    \label{fig:recurrence_network}
\end{figure}

SON models the energy distribution of point cloud through a combination of attractive and repulsive forces acting among neighboring points. Specifically, the 3D object's point cloud is represented as a network, where each node corresponds to a point in the point cloud and edges are formed based on spatial distance. Let $ G = (V, E) $ denote the network model of 3D object, where $ V $ is the set of point cloud points and $ E $ is the set of edges. As illustrated in Figure~\ref{fig:recurrence_network}, an edge is established between two nodes if one lies within the spherical neighborhood of radius $\xi$ centered at the other. The adjacency matrix $ A \in \{0,1\}^{N \times N} $ of the recurrence network is computed as:

\begin{equation}
    \bm{A}_{ij} = H\left( \xi - \| \bm{v}_i - \bm{v}_j \| \right), \quad \bm{v}_i, \bm{v}_j \in V,
\end{equation}
where $ H(\cdot) $ is the Heaviside step function, $ \xi $ is the radius threshold, and $ \bm{v}_i, \bm{v}_j \in \mathbb{R}^3 $ are the coordinates of point cloud points $ i $ and $ j $, respectively. It is defined that $A_{ii} = 0$ to exclude self-connections.

Then the spring-electrical model is employed to model the energy distribution of the point cloud. Network nodes are simulated as electrically charged particles, where the repulsive force between any pair of nodes is computed as:
\begin{equation}
    \text{RF}(i, j) = -\frac{CK^{p+1}}{\|\bm{v}_i - \bm{v}_j\|^p} (\bm{v}_i - \bm{v}_j) , \quad \bm{v}_i \neq \bm{v}_j,\ p > 0,
\end{equation}
where $ C $ is a scaling factor, $ K $ is the natural spring length, and $ p $ controls the decay rate of the repulsive force.

Edges are simulated as springs that pull connected nodes towards each other. The attractive force is given by:
\begin{equation}
    \text{AF}(i, j) = \frac{\|\bm{v}_i - \bm{v}_j\|}{K} (\bm{v}_i - \bm{v}_j), \quad \bm{v}_i \leftrightarrow \bm{v}_j \,,
\end{equation}
where $\bm{v}_i \leftrightarrow \bm{v}_j$ indicates that nodes $i$ and $j$ are connected by an edge, $ K $ is the natural spring length.

The objective is to find the optimal layout by minimizing the total energy of the system, which is defined as:

\begin{equation}
\label{eq:energy}
\mathrm{E}(V, K, C) = \sum_{i \in V} f^2(i, \bm{v}_i, K, C) \,,
\end{equation}

\begin{equation}
\label{eq:force}
\bm{f}(i, \bm{v}_i, K, C) = \sum_{\bm{v}_i \neq \bm{v}_j} -\frac{CK^{p+1}}{\|\bm{v}_i - \bm{v}_j\|^p} (\bm{v}_i - \bm{v}_j) + \sum_{\bm{v}_i \leftrightarrow \bm{v}_j} \frac{\|\bm{v}_i - \bm{v}_j\|}{K} (\bm{v}_i - \bm{v}_j) \,,
\end{equation}
\noindent where $ \bm{f}(i, \bm{v}_i, K, C) $ denotes the combined force acting on node $ i $. This network's energy can be minimized iteratively by updating the positions of the nodes in the direction of the total force vector on them, as shown in Equation~\eqref{eq:update_position}, where $\delta$ is the step size:

\begin{equation}
    \bm{v}_i \leftarrow \bm{v}_i + \delta \times \frac{\bm{f}(i, \bm{v}_i, K, C)}{\|\bm{f}(i, \bm{v}_i, K, C)\|}.
    \label{eq:update_position}
\end{equation}

\subsubsection{\emph{Relationship between SON energy distribution and anomalies}} \label{s:methods.2.2}
This section investigates the relationship between Equation.~\eqref{eq:force} and surface anomalies. Let $\bm{v}_i$ be a point in the point cloud $\bm{V} = [\bm{v}_1, \dots, \bm{v}_N]^\top \in \mathbb{R}^{N \times 3}$, and define its neighborhood set as
\begin{equation}
\mathcal{N}(i) = \left\{ \bm{v}_j \mid \|\bm{v}_j - \bm{v}_i\| < \xi,\ j \ne i \right\},
\end{equation}
where $\xi > 0$ is a given radius parameter.

According to Graph Signal Processing(GSP) theory~\citep{ortegaGraph2018}, if the local neighborhood of each point $\bm{v}_i$ is symmetric around $\bm{v}_i$, the surface of the entire point cloud can be regarded as smooth. In such cases, symmetry implies that:
\begin{equation}
\sum_{\bm{v}_i \neq \bm{v}_j} \frac{\bm{v}_i - \bm{v}_j}{\|\bm{v}_i - \bm{v}_j\|^p} = \vec{0}, \quad \sum_{\bm{v}_i \leftrightarrow \bm{v}_j} \frac{\|\bm{v}_i - \bm{v}_j\|}{K} (\bm{v}_i - \bm{v}_j) = \vec{0}
\quad \Longrightarrow \quad
\bm{f}(i, \bm{v}_i, K, C) = \vec{0} \,.
\end{equation}

\begin{figure}
\centering
\includegraphics[width=0.3\textwidth]{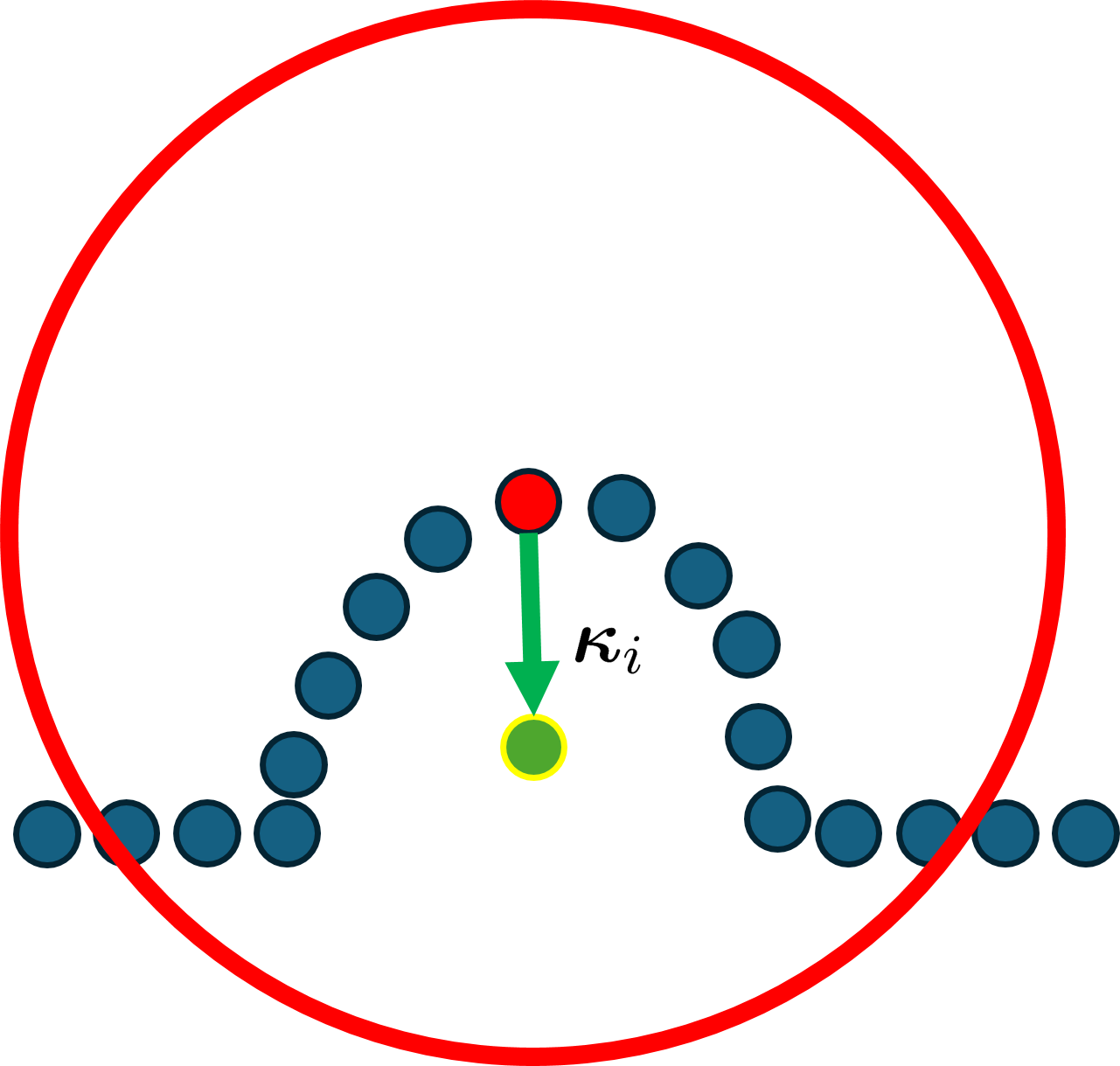}
\caption{Common surface anomaly represented by 3D point clouds: dent}
\label{fig:deviate_from_neighbor}
\end{figure}

As illustrated in Figure~\ref{fig:deviate_from_neighbor}, if the surface of the point cloud contains anomalies such as dents, the local neighborhood of a point $\bm{v}_i$ is not centrally symmetric. We define the local centroid $\bm{c}_i$ of point $\bm{v}_i$,
\begin{equation}
\bm{c}_i = \frac{1}{|\mathcal{N}(i)|} \sum_{j \in \mathcal{N}(i)} \bm{v}_j \,.
\end{equation}

The vector $\bm{\kappa}_i = \bm{v}_i - \bm{c}_i$ captures the offset of the point $\bm{v}_i$ relative to its neighbors and quantifies the degree of local geometric asymmetry. Furthermore, we define the unit direction vector $\bm{n}$ corresponding to $\bm{\kappa}_i$ as
\begin{equation}
\bm{n} = \frac{\bm{\kappa}_i}{\|\bm{\kappa}_i\|} = \frac{\bm{v}_i - \bm{c}_i}{\|\bm{v}_i - \bm{c}_i\|} \,.  
\end{equation}

By employing the proposed unit direction vector $\bm{n}$, we establish a connection between the lower bound of $\|\bm{f}(i)\|$ and $\|\bm{\kappa}_i\|$. Finally, in Theorem~\ref{thm:lower_bound}, we prove that the lower bound of $\|\bm{f}(i)\|$ is proportional to $\|\bm{\kappa}_i\|$.

\begin{theorem}
\label{thm:lower_bound}
It holds that
\begin{equation}
\|\bm{f}(i)\| \geq \|\bm{n} \cdot \bm{f}(i)\| \geq \alpha \|\bm{\kappa}_i\| - \beta,
\end{equation}
where $\alpha$ and $\beta$ are constants once the point cloud is given.
\end{theorem}

The proof is provided in Appendix of the supplementary material. This inequality demonstrates that local geometric asymmetry induces a force component, and the lower bound of the force is proportional to $\|\bm{\kappa}_i\|$. This establishes a direct link between geometric anomalies and the resulting force, thereby providing a theoretical foundation for SONAR. 

\subsubsection{\emph{Energy local normalization and anomaly ranking}} \label{s:methods.2.3}

Based on the sparse anomalies assumption and Theorem~\ref{thm:lower_bound}, the core criterion of SONAR for identifying a point as anomalous is: if a point has higher energy compared to its neighboring points, it is more likely to be anomalous. In other words, points with higher locally normalized energy are considered more anomalous, rather than focusing on points with high absolute energy values.

We choose local Z-score normalization to the computed locally normalized energy, after modeling the absolute energy distribution of point clouds.
\begin{equation}
    \mathrm{E}_{\text{normalized}} = \frac{\mathrm{E} - \mu_\mathrm{E}}{\sigma_\mathrm{E}},
    \label{eq:z_score_energy}
\end{equation}
where \( \mathrm{E} \) denotes the original energy of a point, \( \mu_\mathrm{E} \) is the mean energy in its local region, and \( \sigma_\mathrm{E} \) is the corresponding standard deviation. After local normalization, the normalized energy of points with anomalous tendencies is significantly greater than 0, while the energy of normal points is close to 0 or negative.

To further emphasize the high energy components and ignore the low-energy components, we apply a clipped ReLU activation function to the standardized energy. This operation retains the energy of points with relative energy greater than 0 (absolute energy above the mean), while setting all components with relative energy less than 0 (absolute energy below the mean) to zero. Additionally, it caps the output at a maximum value of 2 to prevent excessively large activations, ensuring that all relative energy values fall within the same range for easier visualization and subsequent ranking. The resulting transformation is given by:

\begin{equation}
\text{Clipped ReLU}(\mathrm{E}_{\text{normalized}}) = \min\left( \max(0, \mathrm{E}_{\text{normalized}}),\ 2 \right).
\label{eq:relu_z_score_energy}
\end{equation}

We identify anomalous points through a two-stage ranking process. In the first stage, we rank points based on the normalized energy after applying the Clipped ReLU function and select the top $\gamma\%$ of high energy points as seed points. In the second stage, the optimizable point set $\mathrm{P}$ is defined as the union of the neighborhoods of all seed points, while the remaining points remain fixed. The positions of points in $\mathrm{P}$ are then updated according to Equation~\eqref{eq:update_position}. Points are re-ranked by descending displacement magnitude, and those falling within the top $\delta\%$ and exceeding a distance threshold $\tau$ are identified as anomaly points. The first threshold ($\delta\%$) enforces the sparse anomaly assumption, while the second threshold ($\tau$) primarily aims to exclude minor displacements caused by noise, non-uniform point cloud density, and similar artifacts.

\subsubsection{\emph{Energy boundary condition for open surface}} \label{s:methods.2.4}

\begin{figure}[htbp]
\centering
\begin{subfigure}{0.35\textwidth}
  \centering
  \includegraphics[width=\textwidth]{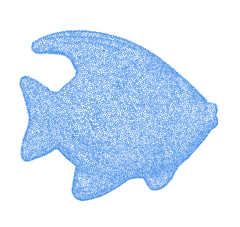}
  \caption{Original point cloud of fish}
  \label{fig:fish_raw}
\end{subfigure}
\hfill
\begin{subfigure}{0.35\textwidth}
  \centering
  \includegraphics[width=\textwidth]{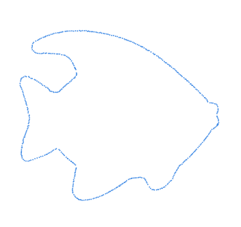}
  \caption{Point cloud boundary of fish}
  \label{fig:fish_boundary}
\end{subfigure}
\caption{Energy boundary condition. To avoid false positives at the boundary, the energy of boundary points in Panel (b) is set to a specific value, ensuring that their normalized energy is 0.25.}
\label{fig:boundary_condition_of_energy}
\end{figure}

Most of previous approaches that rely on local geometric information(e.g., normal vectors and curvatures)~\citep{miaoPipeline2022,jovancevic3D2017} tend to flag boundary points of open surfaces as anomalies. Due to the lack of symmetric neighborhood structures around boundary points, energy of boundary points are often distorted. Specifically, the absence of neighboring points on one side leads to unbalanced force contributions, which increases the likelihood of false positives. According to the energy local normalization described in Section~\ref{s:methods.2.3}, the boundary energy anomaly issue is partially mitigated, yet some false detections still persist. To address this issue, we introduce energy boundary conditions for open-surface point clouds. As shown in Figure~\ref{fig:boundary_condition_of_energy}, we use the Angle Criterion (AC)~\citep{Rusu_ICRA2011_PCL} method to extract boundary points from the point cloud. As shown in Equation~\eqref{eq:boundary_condition}, we assign boundary points and their neighborhoods a fixed normalized energy value of 0.25. This approach suppresses the anomaly detection issue at the boundaries while not affecting the detection outside the boundary.

\begin{equation}
    \mathrm{E}_{\text{normalized}}(\text{boundaries}) = 0.25
    \label{eq:boundary_condition}
\end{equation}

\section{Case study} \label{s:case study}
In this section, we conduct numerical and case studies to evaluate SONAR's performance. Section~\ref{s:case study 1} compares the proposed SONAR with several benchmark methods under comprehensive numeral settings. Furthermore, in Section ~\ref{s:case study 2}, more samples are provided to validate the feasibility of the proposed SONAR in real-world applications. All computational processes were carried out on a desktop computer equipped with an Intel Core i7-13700 CPU and an NVIDIA GeForce RTX 4070 Super GPU with 12 GB GDDR6X memory.

\subsection{\emph{Experimental results}} \label{s:case study 1}

\subsubsection{\emph{Data description and parameter setting}} \label{s:case study 1.1}
To evaluate SONAR's of the proposed SONAR, we simulate the hole-type anomaly by constructing two models: a cube (edge length: 64) with multiple conical protrusions (radius: 4, height: 10), and a triangular pyramid (edge length: 64) with the same type of protrusions. For each point, we add Gaussian white noise with a variance of 0.1 in all three XYZ directions. 

In the numerical case study, to accelerate computation and eliminate the influence of varying input sizes, each point cloud is first downsampled using Farthest Point Sampling (FPS)~\citep{9919246} to about 10k points and rescaled into a $64 \times 64 \times 64$ cube. We set the parameters defined in Section~\ref{s:methods.2.3} as follows: $\gamma = 0.03$ to select the highest-energy points as seed points, and $\delta = 5$ to satisfy the sparsity assumption of anomalies. The distance threshold $\tau$ is set as the average of the nearest neighbor distances among points in the point cloud, thereby excluding points with negligible absolute displacement.

\subsubsection{\emph{Baselines}} \label{s:case study 1.2}
We compare the proposed SONAR method with the following benchmark methods:
\begin{enumerate}
    \item Statistical Outlier Removal (SOR)-based method ~\citep{baltaFast2018}: The method detects anomalies by checking the average distance between each point and its neighbors. Points with much larger distances than expected are considered anomalies.
    \item FPFH-based method~\citep{miaoPipeline2022}: Conventional methods use the FPFH descriptor to extract local geometric features and identify anomaly points based on the Isolation Forest~\citep{liuIsolation2008}.
    \item PointSGRADE~\citep{taoPointSGRADESparseLearning2025}: This method represents the point cloud as a graph, assumes surface smoothness, and formulates anomaly detection as a sparse optimization problem solved via a majorization-minimization algorithm.
    \item Region growing (RG)-based method~\citep{jovancevic3D2017}: It first denoises the point cloud and then uses the normals and curvatures to merge points into a set of homogeneous regions. All regions except the largest one are regarded as anomalous.
\end{enumerate}

FPFH and RG-based methods are commonly used approaches that rely on local geometric features (normals and curvatures). In contrast, PointSGRADE is a representative method based on global geometric features. While SOR-based methods are often designed for noise reduction, the detected outliers in these studies correspond to anomalous points deviating from typical patterns. By selecting these representative methods for comparison, we ensure that the pipeline covers a wide range of strategies, thereby enabling a comprehensive assessment of the proposed method.

\subsubsection{\emph{Quantitative indices}} \label{s:case study 1.3}
We denote $\mathrm{FP}$ and $\mathrm{TN}$ as the number of anomaly points predicted incorrectly and correctly, respectively. Similarly, $\mathrm{FN}$ and $\mathrm{TP}$ represent the number of reference surface points predicted erroneously and correctly, respectively. To comprehensively evaluate the performance of anomaly detection, we adopt four metrics:

\begin{enumerate}
    \item Precision:
    \[
    \mathrm{Precision} = \frac{\mathrm{TP}}{\mathrm{TP} + \mathrm{FP}}
    \]
    measuring the proportion of detected anomalies that are actually true anomalies.

    \item Recall:
    \[
    \mathrm{Recall} = \frac{\mathrm{TP}}{\mathrm{TP} + \mathrm{FN}}
    \]
    indicating the proportion of actual anomalies that are correctly identified.

    \item F1-score:
    \[
    \mathrm{F1\text{-}score} = 2 \cdot \frac{\mathrm{Precision} \cdot \mathrm{Recall}}{\mathrm{Precision} + \mathrm{Recall}}
    \]
    representing the harmonic mean of Precision and Recall, providing a balanced measure of both.

    \item AUROC:
    \[
    \mathrm{AUROC} = \int_{0}^{1} \mathrm{TPR}(FPR^{-1}(x)) \, dx,
    \]
    where
    \begin{align*}
    \mathrm{FPR} &= \frac{\mathrm{FP}}{\mathrm{FP} + \mathrm{TN}}, \\
    \mathrm{TPR} &= \frac{\mathrm{TP}}{\mathrm{TP} + \mathrm{FN}}.
    \end{align*}
    AUROC denotes the area under the receiver operating characteristic curve, which evaluates the model's ability to distinguish between normal and anomalous points across different threshold settings.
\end{enumerate}

Higher values of these metrics reflect better detection performance.

\subsubsection{\emph{Results on representative numerical samples}} \label{s:case study 1.4}
\begin{figure}[htbp]
    \centering
    \begin{subfigure}[b]{0.25\linewidth}
        \centering
        \includegraphics[width=\linewidth]{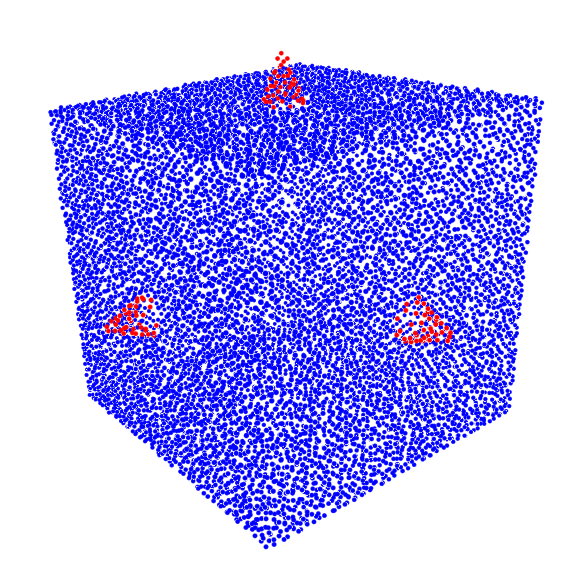}
        \caption{Ground truth}
        \label{fig:subfig1-1}
    \end{subfigure}%
    \hfill
    \begin{subfigure}[b]{0.25\linewidth}
        \centering
        \includegraphics[width=\linewidth]{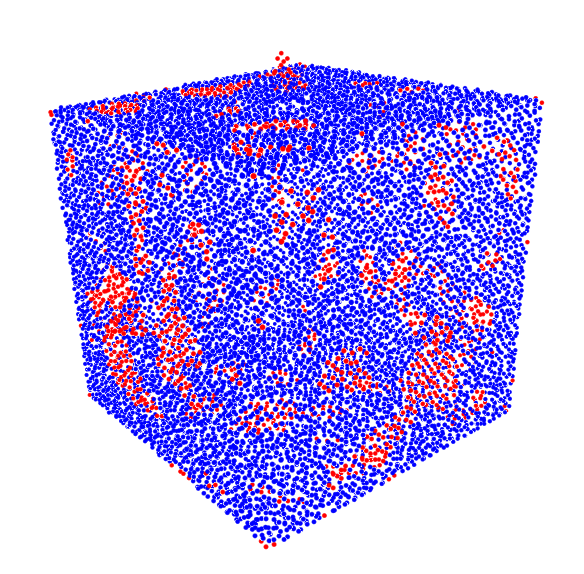}
        \caption{SOR-based method}
        \label{fig:subfig1-2}
    \end{subfigure}%
    \hfill
    \begin{subfigure}[b]{0.25\linewidth}
        \centering
        \includegraphics[width=\linewidth]{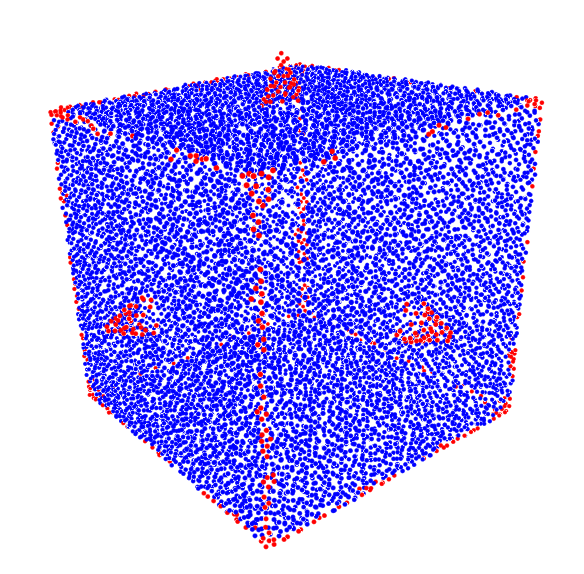}
        \caption{FPFH-based method}
        \label{fig:subfig1-3}
    \end{subfigure}
    \vspace{1em}
    
    \begin{subfigure}[b]{0.25\linewidth}
        \centering
        \includegraphics[width=\linewidth]{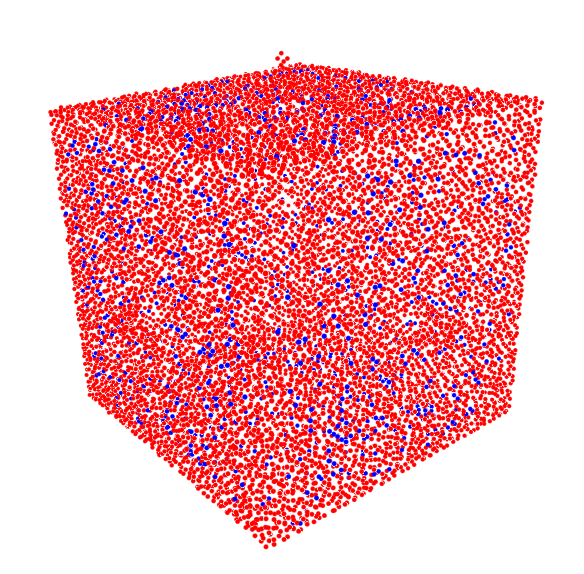}
        \caption{PointSGRADE}
        \label{fig:subfig1-4}
    \end{subfigure}%
    \hfill
    \begin{subfigure}[b]{0.25\linewidth}
        \centering
        \includegraphics[width=\linewidth]{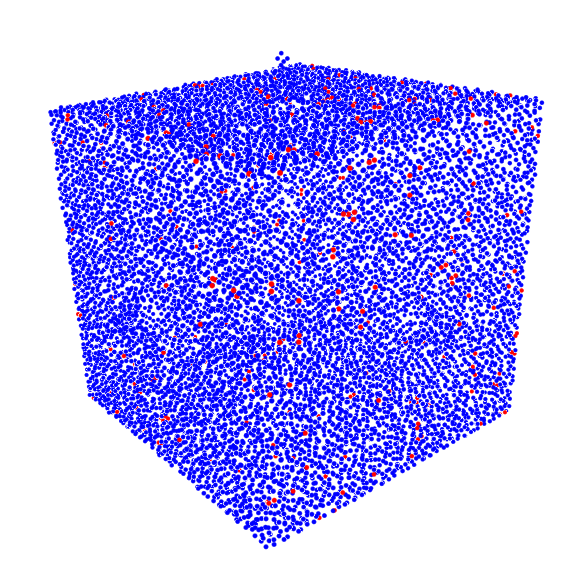}
        \caption{RG-based method}
        \label{fig:subfig1-5}
    \end{subfigure}%
    \hfill
    \begin{subfigure}[b]{0.25\linewidth}
        \centering
        \includegraphics[width=\linewidth]{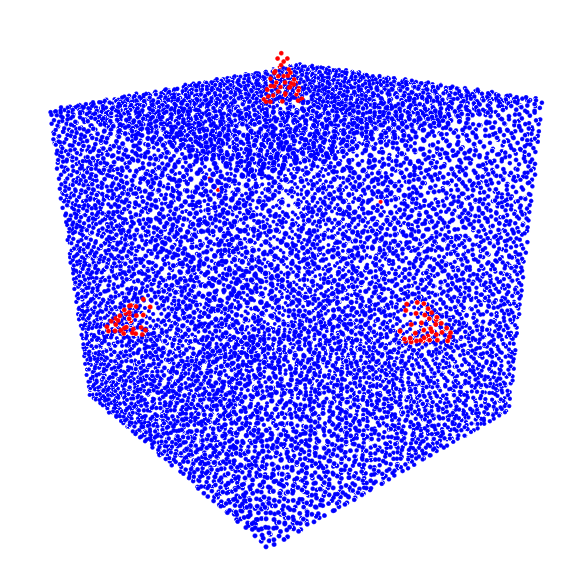}
        \caption{SONAR}
        \label{fig:subfig1-6}
    \end{subfigure}
    
    \caption{Anomaly detection of a closed cube with a conical protrusion}
    \label{fig:anomaly_results1}
\end{figure}

\begin{figure}[htbp]
    \centering
    \begin{subfigure}[b]{0.25\linewidth}
        \centering
        \includegraphics[width=\linewidth]{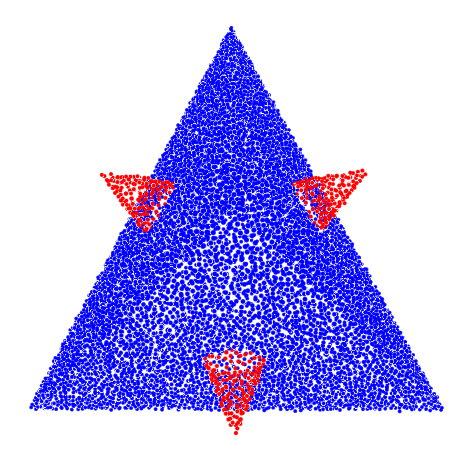}
        \caption{Ground truth}
        \label{fig:subfig2-1}
    \end{subfigure}%
    \hfill
    \begin{subfigure}[b]{0.25\linewidth}
        \centering
        \includegraphics[width=\linewidth]{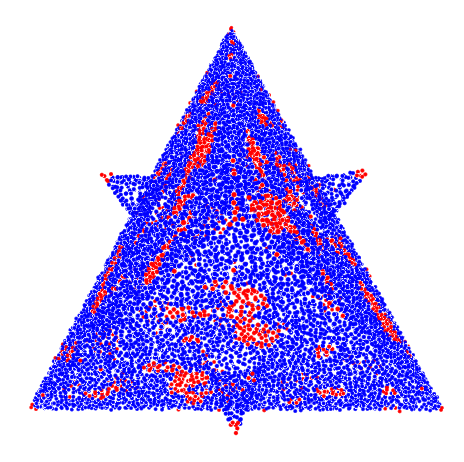}
        \caption{SOR-based method}
        \label{fig:subfig2-2}
    \end{subfigure}%
    \hfill
    \begin{subfigure}[b]{0.25\linewidth}
        \centering
        \includegraphics[width=\linewidth]{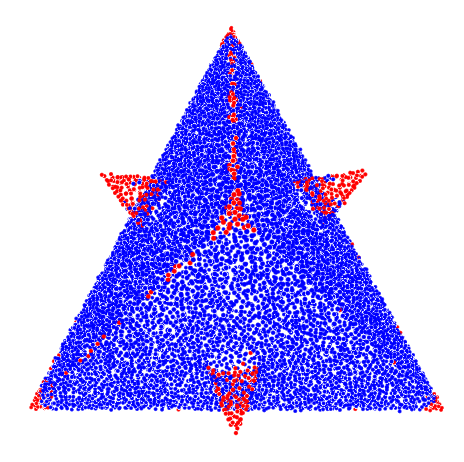}
        \caption{FPFH-based method}
        \label{fig:subfig2-3}
    \end{subfigure}
    \vspace{1em}
    
    \begin{subfigure}[b]{0.25\linewidth}
        \centering
        \includegraphics[width=\linewidth]{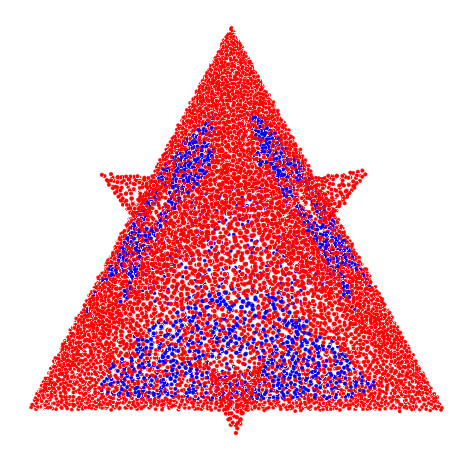}
        \caption{PointSGRADE}
        \label{fig:subfig2-4}
    \end{subfigure}%
    \hfill
    \begin{subfigure}[b]{0.25\linewidth}
        \centering
        \includegraphics[width=\linewidth]{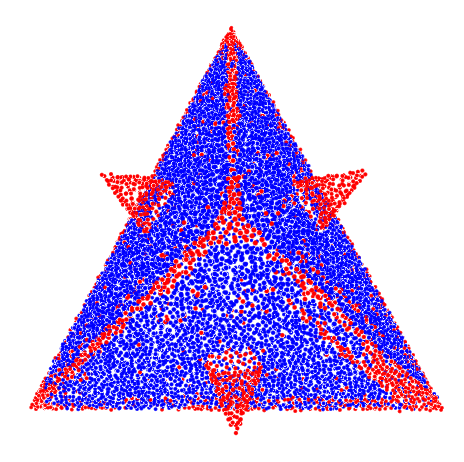}
        \caption{RG-based method}
        \label{fig:subfig2-5}
    \end{subfigure}%
    \hfill
    \begin{subfigure}[b]{0.25\linewidth}
        \centering
        \includegraphics[width=\linewidth]{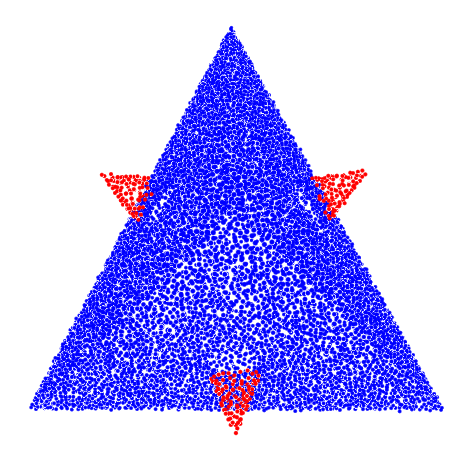}
        \caption{SONAR}
        \label{fig:subfig2-6}
    \end{subfigure}
    
    \caption{Anomaly detection of a closed tetrahedron with a conical protrusion}
    \label{fig:anomaly_results2}
\end{figure}

This section validates the performance of the proposed SONAR method by conducting comparative experiments with the two datasets from section~\ref{s:case study 1.1} and the baselines from section~\ref{s:case study 1.2}. As shown in Figures~\ref{fig:anomaly_results1}–\ref{fig:anomaly_results2}, the anomaly detection results of SONAR and several baseline methods are visualized, with blue points representing normal regions and red points indicating anomalies. Table~\ref{tab:comparison_results1} compares their performance using common metrics: Precision, Recall, F1-score, and AUROC. SONAR achieves the highest Precision, F1-score, and AUROC among all methods. This shows that it effectively reduces false alarms while maintaining balanced detection performance. SONAR’s Recall and FPFH are similar to each other but notably lower than those of PointsGRADE. The reason is simple: PointsGRADE labels almost all points as anomalies. This leads to a very high recall but also causes many false positives.

\begin{figure}[htbp]
    \centering
    \begin{subfigure}[b]{0.3\linewidth}
        \centering
        \includegraphics[width=\linewidth]{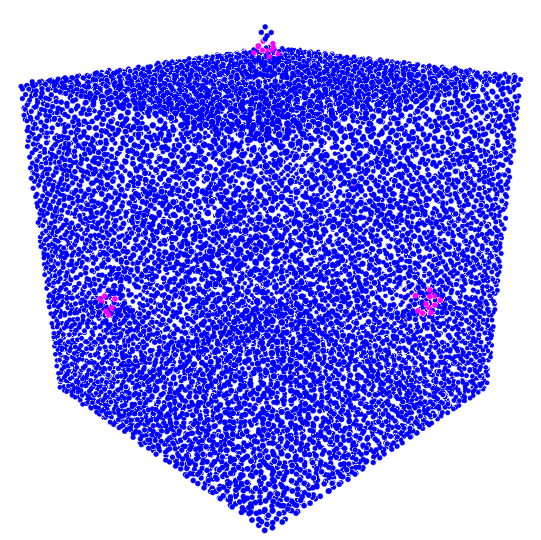}
        \caption{High energy seed points}
        \label{fig:ranking_visualization1-1}
    \end{subfigure}
    \hfill
    \begin{subfigure}[b]{0.3\linewidth}
        \centering
        \includegraphics[width=\linewidth]{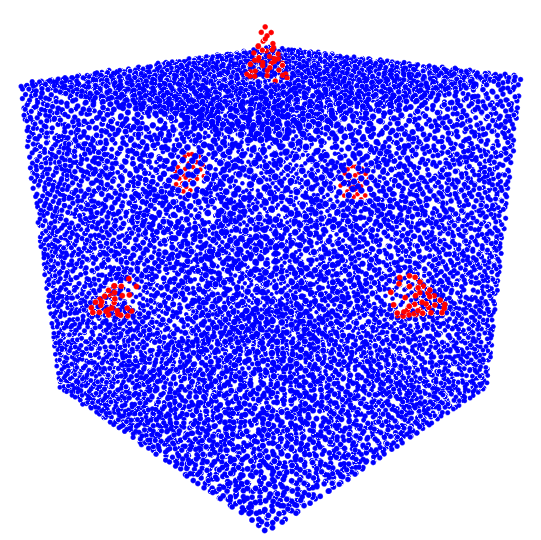}
        \caption{Optimizable point set}
        \label{fig:ranking_visualization1-2}
    \end{subfigure}
    \hfill
    \begin{subfigure}[b]{0.3\linewidth}
        \centering
        \includegraphics[width=\linewidth]{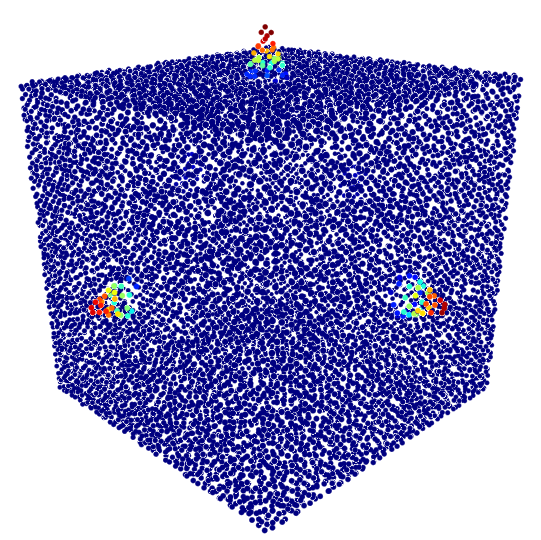}
        \caption{Displacement of the points}
        \label{fig:ranking_visualization1-3}
    \end{subfigure}
    
    \begin{subfigure}[b]{\linewidth}
        \centering
        \includegraphics[width=\linewidth]{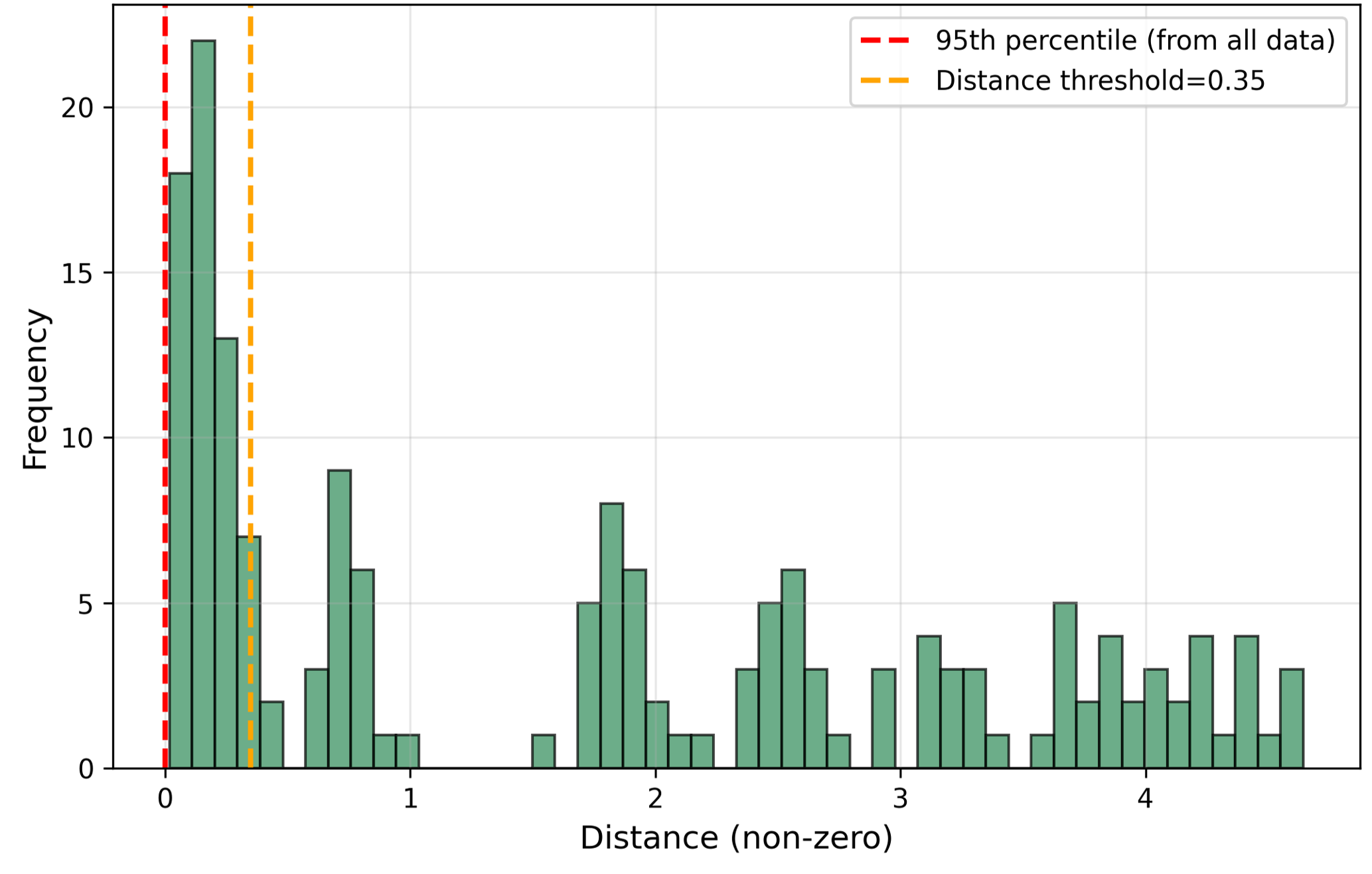}
        \caption{Histogram of point displacements (excluding zero displacements)}
        \label{fig:ranking_visualization1-4}
    \end{subfigure}
    
    \caption{Ranking of point displacements}
    \label{fig:ranking_visualization_1}
\end{figure}

As shown in Figure~\ref{fig:ranking_visualization_1}, the anomaly ranking process is illustrated. In Figure~\ref{fig:ranking_visualization_1}(\subref{fig:ranking_visualization1-1}), the red points denote the selected high energy seed points (the top $\gamma = 0.03\%$ of high energy points). Figure~\ref{fig:ranking_visualization_1}(\subref{fig:ranking_visualization1-2}) displays the optimizable point set $\mathrm{P}$, consisting of the high energy seed points and their neighborhoods. Through the force-directed optimization process defined in Equation~\eqref{eq:update_position}, these points are displaced. Figure~\ref{fig:ranking_visualization_1}(\subref{fig:ranking_visualization1-3}) shows the displacement magnitudes of all points, where warmer colors (closer to red) indicate larger displacements and cooler colors (closer to blue) indicate smaller displacements. Figure~\ref{fig:ranking_visualization_1}(\subref{fig:ranking_visualization1-4}) presents the histogram of point displacements. Since the majority of points have zero displacement, we exclude them from visualization and only display points with displacement greater than zero. The top 95\% (top $\delta$\%) and the distance threshold $\tau = 0.35$ are annotated on the plot. Only points whose displacements exceed both thresholds are classified as anomalies, and the resulting anomaly detection is shown in Figure~\ref{fig:anomaly_results1}(\subref{fig:subfig1-6}).


\begin{table}[htbp]
\centering
\caption{Results of the numerical case study}
\label{tab:comparison_results1}
\begin{tabular}{lcccc}
\hline
Method $\backslash$ Metric & Precision & Recall & F1-score & AUROC \\
\hline
SOR & 0.0349 & 0.2512 & 0.0581 & 0.5583 \\
FPFH & 0.4238 & 0.7543 & 0.4934 & \textbf{0.8624} \\
PointSGRADE & 0.0351 & \textbf{0.9602} & 0.0668 & 0.5483 \\
RG & 0.1189 & 0.5004 & 0.1916 & 0.7000 \\
SONAR & \textbf{0.9867} & 0.7171 & \textbf{0.8249} & \textbf{0.8585} \\
\hline
\end{tabular}
\end{table}

\subsection{\emph{Real case study}} \label{s:case study 2}

\begin{figure}[htbp]
    \centering
    \begin{subfigure}[b]{0.25\linewidth}
        \centering
        \includegraphics[width=\linewidth]{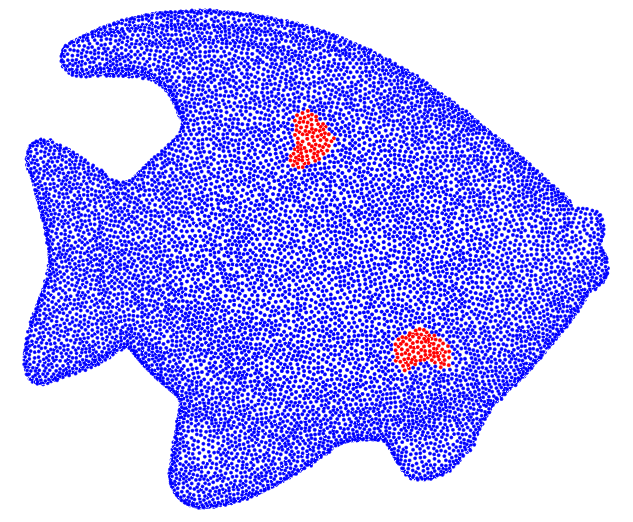}
        \caption{Ground truth}
        \label{fig:subfig3-1}
    \end{subfigure}%
    \hfill
    \begin{subfigure}[b]{0.25\linewidth}
        \centering
        \includegraphics[width=\linewidth]{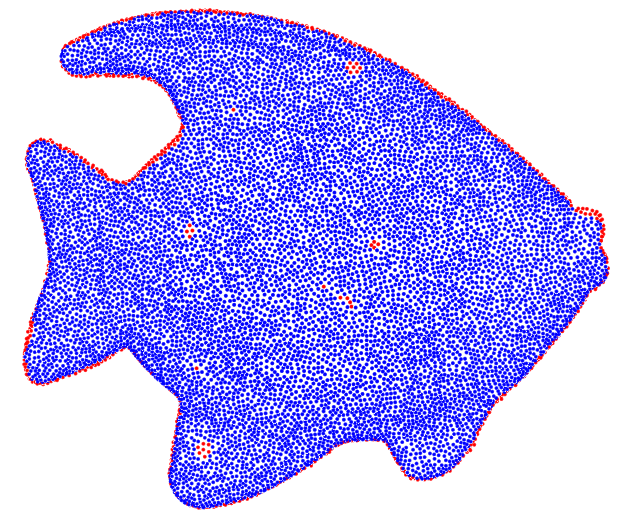}
        \caption{SOR-based method}
        \label{fig:subfig3-2}
    \end{subfigure}%
    \hfill
    \begin{subfigure}[b]{0.25\linewidth}
        \centering
        \includegraphics[width=\linewidth]{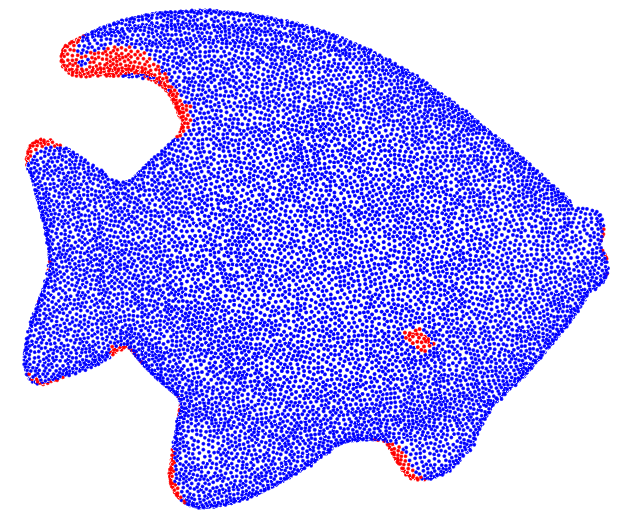}
        \caption{FPFH-based method}
        \label{fig:subfig3-3}
    \end{subfigure}
    \vspace{1em}
    \begin{subfigure}[b]{0.25\linewidth}
        \centering
        \includegraphics[width=\linewidth]{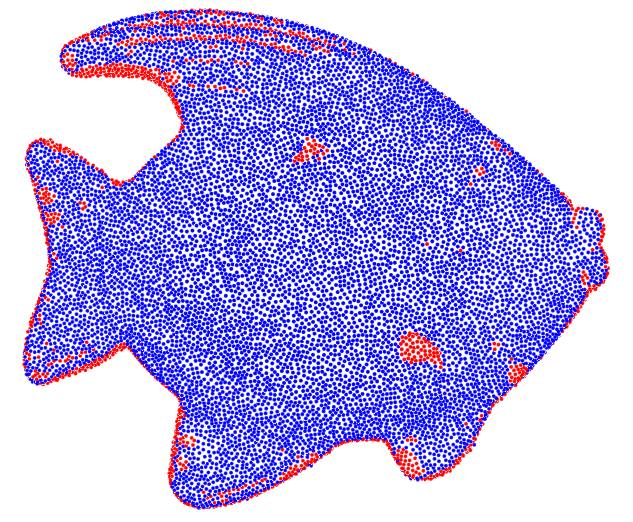}
        \caption{PointSGRADE}
        \label{fig:subfig3-4}
    \end{subfigure}%
    \hfill
    \begin{subfigure}[b]{0.25\linewidth}
        \centering
        \includegraphics[width=\linewidth]{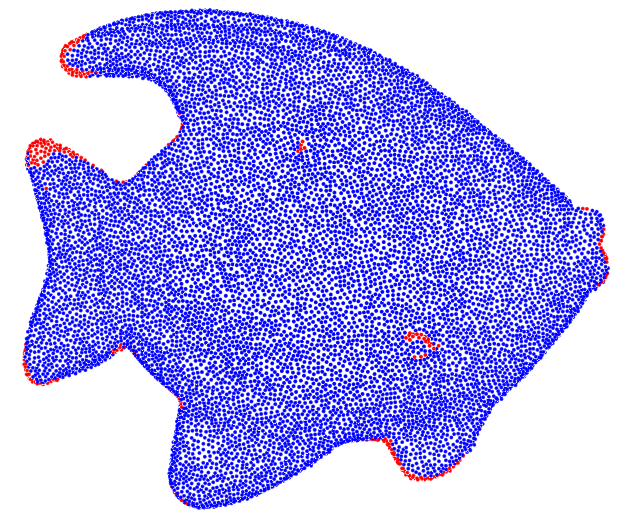}
        \caption{RG-based method}
        \label{fig:subfig3-5}
    \end{subfigure}%
    \hfill
    \begin{subfigure}[b]{0.25\linewidth}
        \centering
        \includegraphics[width=\linewidth]{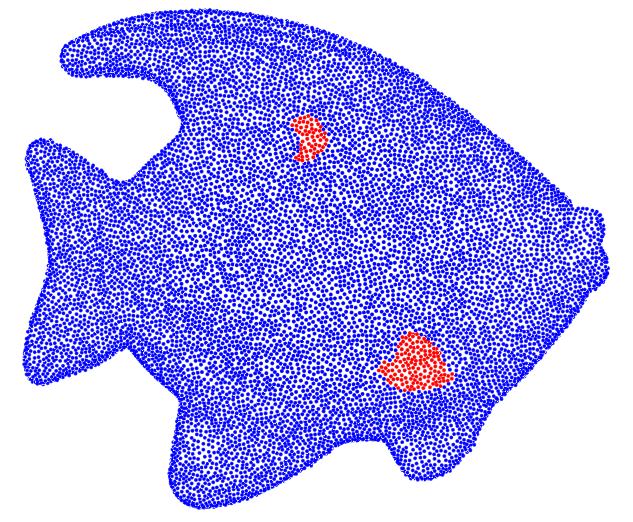}
        \caption{SONAR}
        \label{fig:subfig3-6}
    \end{subfigure}
    
    \caption{Anomaly detection of open fish surfaces}
    \label{fig:anomaly_results3}
\end{figure}
\begin{figure}[htbp]
    \centering
    \begin{subfigure}[b]{0.3\linewidth}
        \centering
        \includegraphics[width=\linewidth]{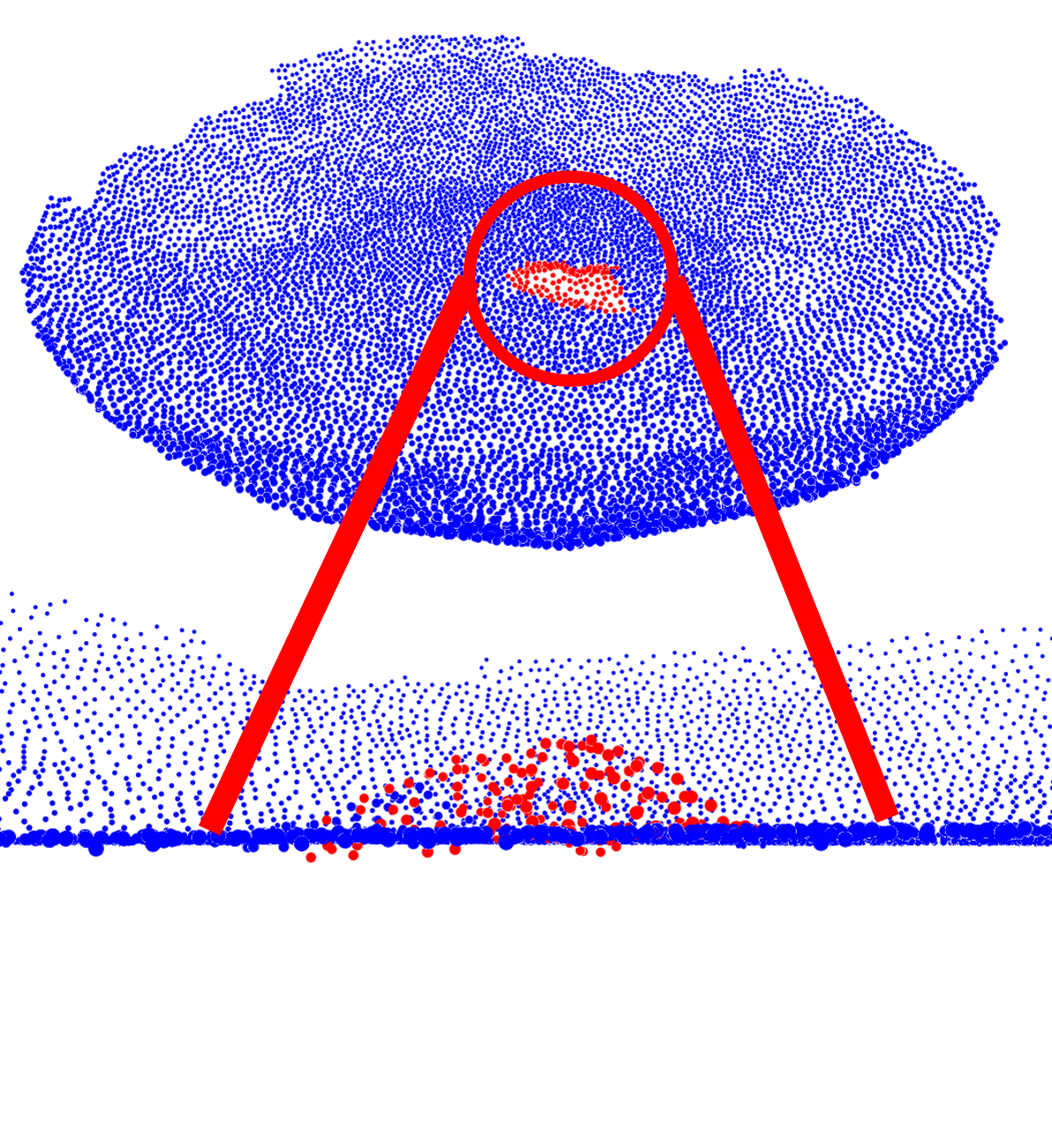}
        \caption{Ground truth}
        \label{fig:subfig4-1-1}
    \end{subfigure}%
    \hfill
    \begin{subfigure}[b]{0.3\linewidth}
        \centering
        \includegraphics[width=\linewidth]{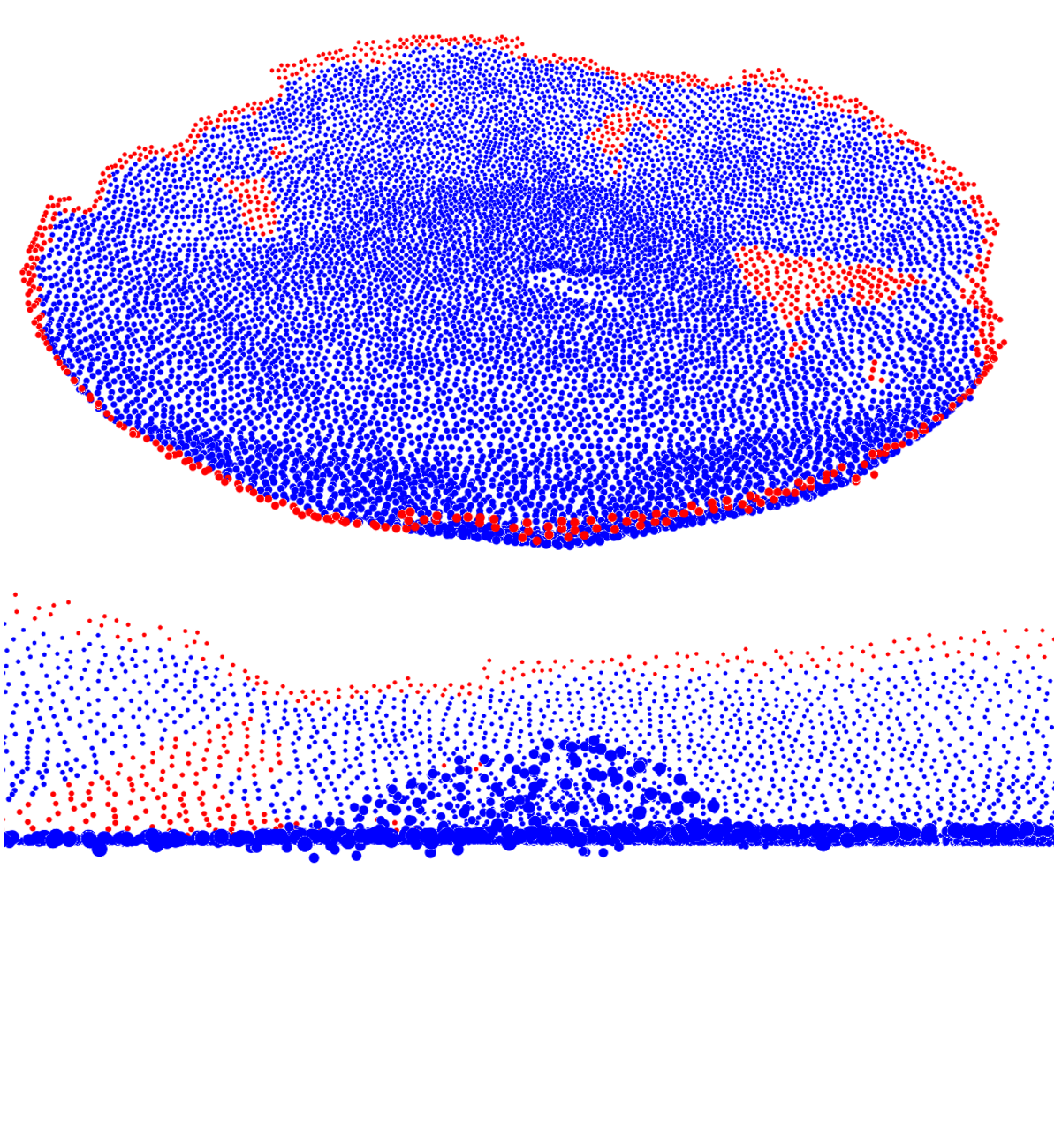}
        \caption{SOR-based method}
        \label{fig:subfig4-2-1}
    \end{subfigure}%
    \hfill
    \begin{subfigure}[b]{0.3\linewidth}
        \centering
        \includegraphics[width=\linewidth]{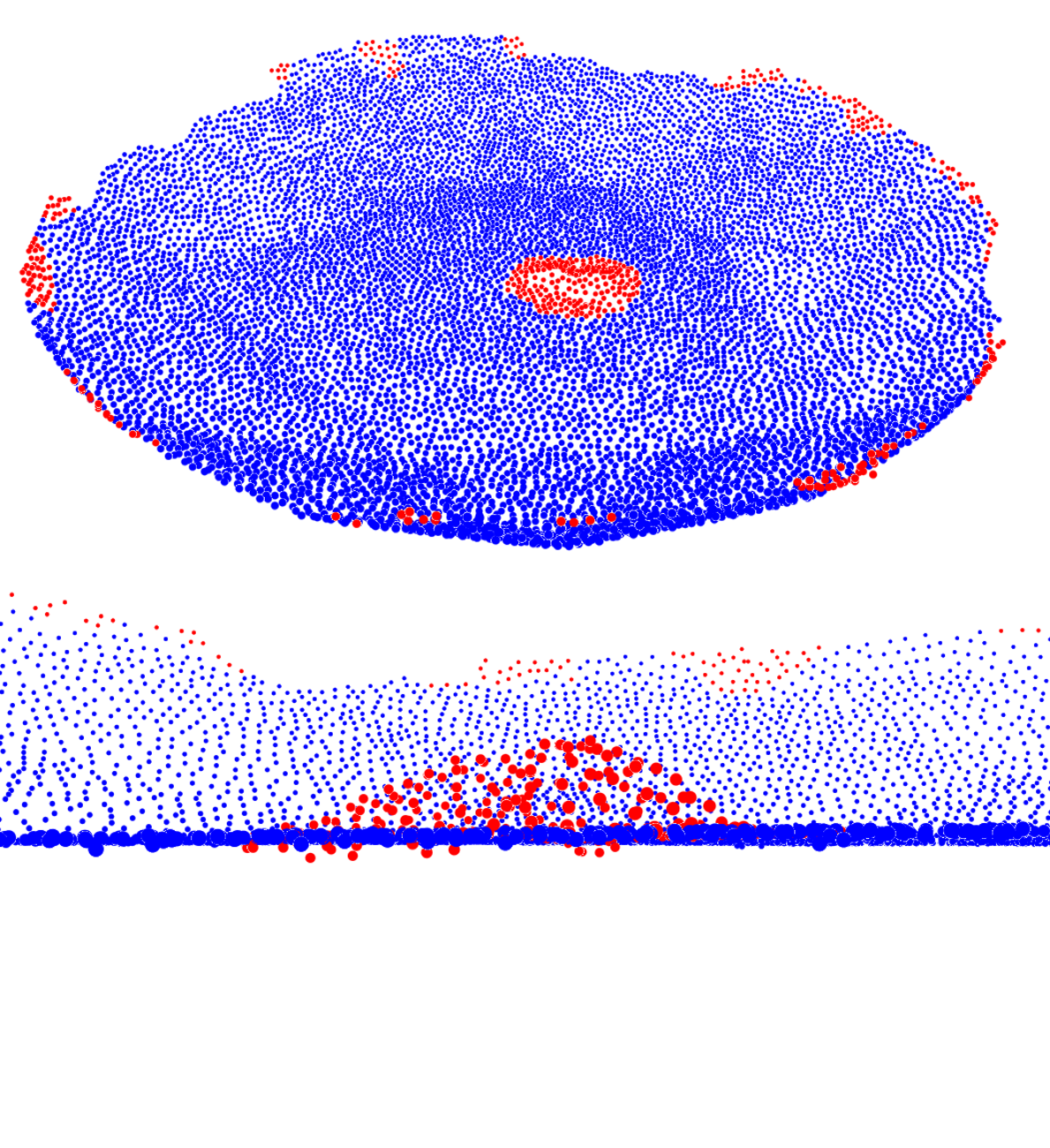}
        \caption{FPFH-based method}
        \label{fig:subfig4-3-1}
    \end{subfigure}
    \vspace{1em}
    \begin{subfigure}[b]{0.3\linewidth}
        \centering
        \includegraphics[width=\linewidth]{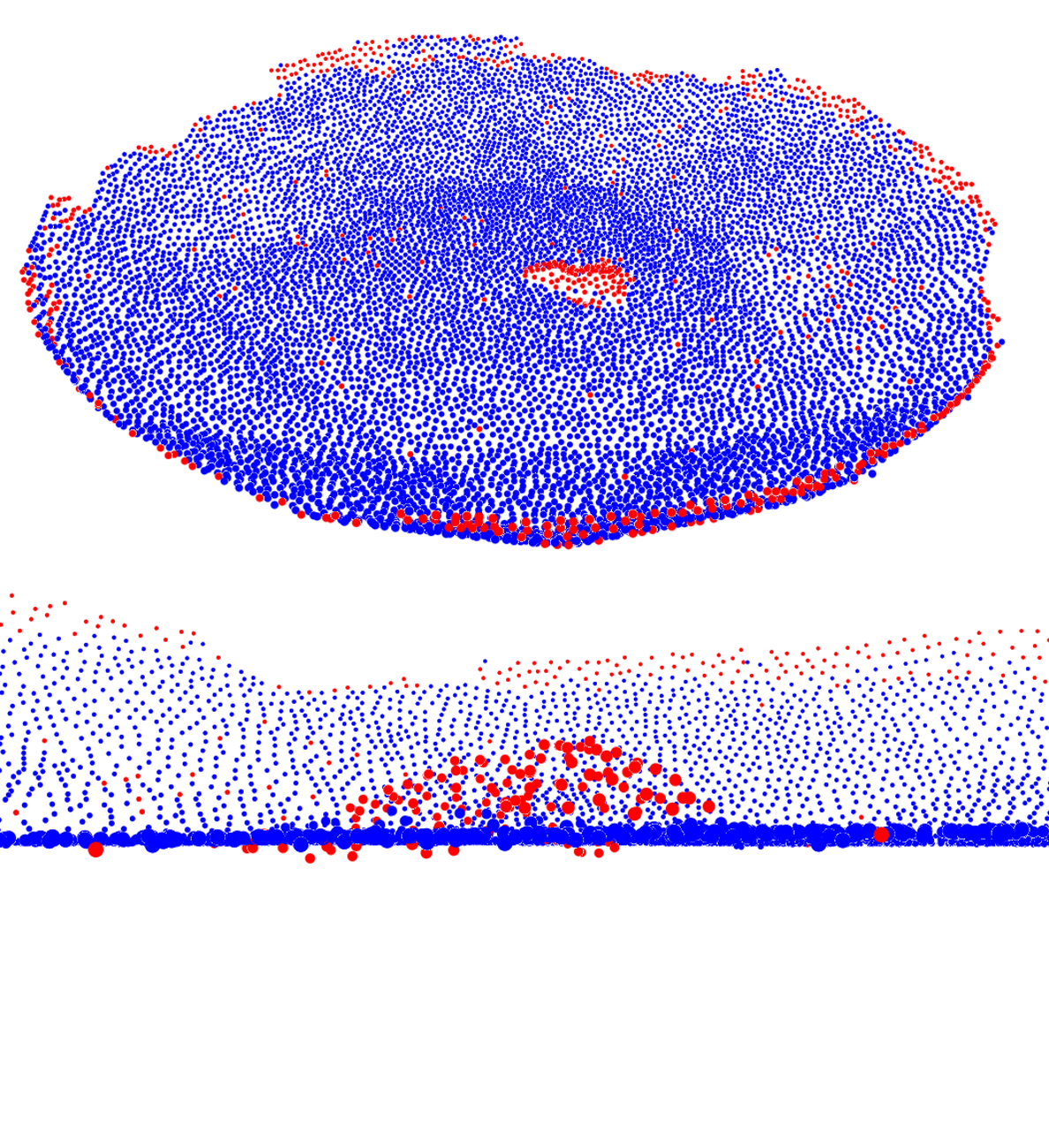}
        \caption{PointSGRADE}
        \label{fig:subfig4-4-1}
    \end{subfigure}%
    \hfill
    \begin{subfigure}[b]{0.3\linewidth}
        \centering
        \includegraphics[width=\linewidth]{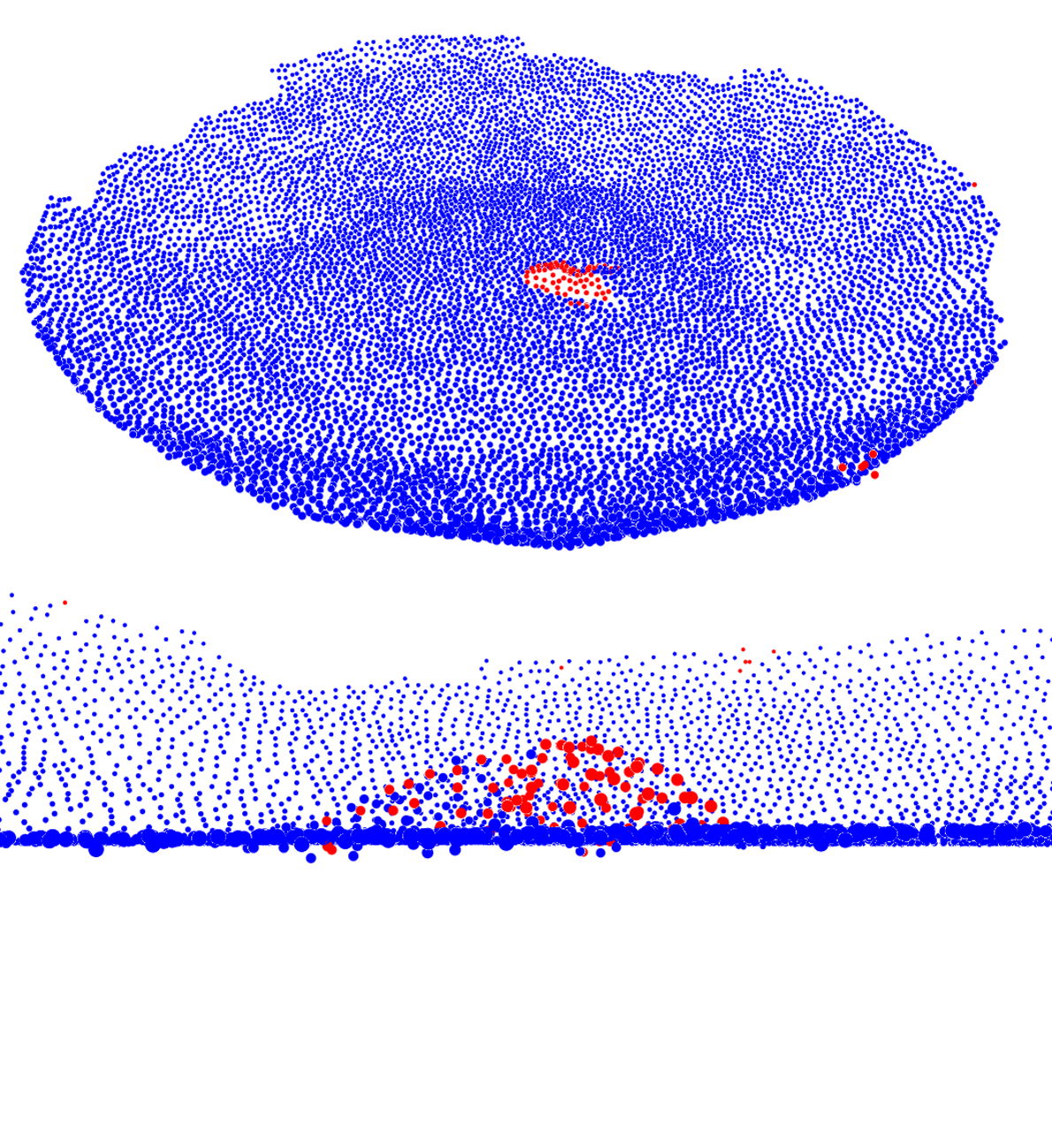}
        \caption{RG-based method}
        \label{fig:subfig4-5-1}
    \end{subfigure}%
    \hfill
    \begin{subfigure}[b]{0.3\linewidth}
        \centering
        \includegraphics[width=\linewidth]{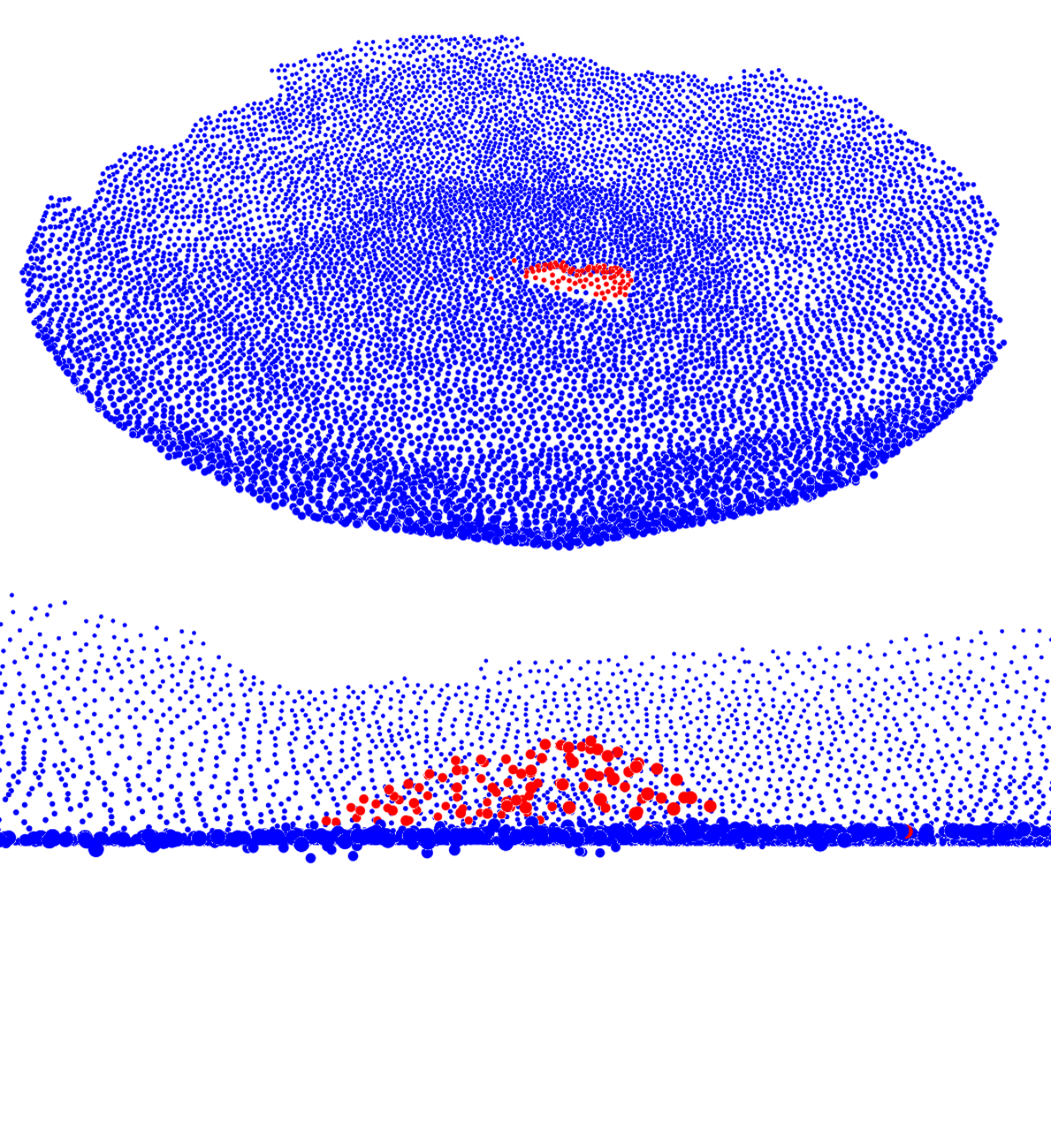}
        \caption{SONAR}
        \label{fig:subfig4-6-1}
    \end{subfigure}
    
    \caption{Anomaly detection of an open diamond: above, global view; below, local view}
    \label{fig:anomaly_results4}
\end{figure}

\begin{table}[htbp]
\centering
\caption{Results of the real case samples}
\label{tab:comparison_results2}
\begin{tabular}{lcccc}
\hline
Method $\backslash$ Metric & Precision & Recall & F1-score & AUROC \\
\hline
SOR & 0.0000 & 0.0000 & 0.0000 & 0.4587 \\
FPFH & 0.1836 & \textbf{0.5604} & 0.2765 & \textbf{0.7594} \\
PointSGRADE & 0.1106 & 0.5480 & 0.1817 & 0.7261 \\
RG & 0.4959 & 0.3419 & 0.3902 & 0.6626 \\
SON & \textbf{0.7299} & \textbf{0.6067} & \textbf{0.6500} & \textbf{0.8010} \\
\hline
\end{tabular}
\end{table}

To validate the performance of the proposed SONAR in real-world applications, two real-world samples (fish and diamond) are obtained from the Real3D-AD~\citep{liuReal3DAD2023}. All preprocessing methods and parameters are consistent with those described in Section~\ref{s:case study 1.1}.

The results are shown in Figure~\ref{fig:anomaly_results3} and Figure~\ref{fig:anomaly_results4}. The proposed SONAR method yields superior performance in the identification of surface anomalies. However, as shown in Figure~\ref{fig:anomaly_results3}, due to abrupt geometric changes at the boundaries of open surfaces, many state-of-the-art methods erroneously classify boundary points as anomalies. Because of the boundary conditions on energy introduced in  Section~\ref{s:methods.2.4}, SONAR effectively avoids false detections at the boundaries. Moreover, as shown in Table~\ref{tab:comparison_results2}, quantitative metrics indicate that SONAR achieves the best performance in all indices.

\section{Conclusion} \label{s:conclusion}
This paper presents SONAR, a novel untrained method for anomaly detection on piecewise smooth surfaces. Built upon SON, the method models the structure as a spring-repulsion dynamical system through topological relationships, enabling the identification of surface anomalies. It can inference directly on single testing point cloud sample without requiring prior training.

This work offers a novel perspective on untrained anomaly detection and highlights the potential of physics-inspired models in the field of industrial inspection tasks where data are scarce. Unlike conventional untrained methods that heavily rely on normal vectors, SONAR does not require normal vector estimation. By incorporating local energy normalization and energy boundary conditions, it improves anomaly detection performance in high-curvatures and non-smooth regions as well as at open boundaries.

Regarding future work, we aim to explore the anomaly repair capability of SONAR. Detecting anomalies is only the first step; understanding how to repair these anomalies is worth future investigation. Moreover, the SONAR method is not limited to surface point clouds, but can also be applied to volumetric structural models (e.g., bone structures derived from CT scans), where the generalization to other modalities and applications is also worth future research.

\if1\blind{
\section*{Acknowledgements}
This work was supported by the [Funding Agency 1] under Grant [number xxxx]; [Funding Agency 2] under Grant [number xxxx]; and [Funding Agency 3] under Grant [number xxxx].} \fi

\bibliographystyle{chicago}
\spacingset{1}
\bibliography{IISE-Trans}

\appendix
\section*{Appendix}
The following provides a detailed proof of the inequality:
\begin{equation}
\|\bm{f}(i)\| \geq \|\bm{n} \cdot \bm{f}(i)\| \geq \alpha \|\bm{\kappa}_i\| - \beta,
\end{equation}
where $\bm{f}(i)$ denotes the net force at point $\bm{v}_i$, $\bm{n}$ is a unit vector derived from local geometry, $\bm{\kappa}_i$ quantifies local geometric asymmetry, and $\alpha$, $\beta$ are parameters that capture sensitivity and suppression effects, respectively. 

We define the local centroid of point $\bm{v}_i$ as the average position of its neighbors:
\begin{equation}
\bm{c}_i = \frac{1}{|\mathcal{N}(i)|} \sum_{j \in \mathcal{N}(i)} \bm{v}_j.
\end{equation}

The vector $\bm{\kappa}_i = \bm{v}_i - \bm{c}_i$ captures the average signed displacement of the point from its neighbors and quantifies the degree of local geometric asymmetry around $\bm{v}_i$. The unit direction vector is defined as:
\begin{equation}
\bm{n} = \frac{\bm{v}_i - \bm{c}_i}{\|\bm{v}_i - \bm{c}_i\|}.
\end{equation}

Starting from the sum over the local neighborhood, we derive the identity:
\begin{equation}
\sum_{j \in \mathcal{N}(i)} \bm{n} \cdot (\bm{v}_i - \bm{v}_j)
= \bm{n} \cdot \left( \sum_{j \in \mathcal{N}(i)} (\bm{v}_i - \bm{v}_j) \right)
= \bm{n} \cdot \left( |\mathcal{N}(i)| \cdot (\bm{v}_i - \bm{c}_i) \right).
\label{eq:neighborhood_sum}
\end{equation}

Since $\bm{n}$ is aligned with $(\bm{v}_i - \bm{c}_i)$, we compute:
\begin{equation}
\bm{n} \cdot (\bm{v}_i - \bm{c}_i)
= \frac{(\bm{v}_i - \bm{c}_i) \cdot (\bm{v}_i - \bm{c}_i)}{\|\bm{v}_i - \bm{c}_i\|}
= \frac{\|\bm{v}_i - \bm{c}_i\|^2}{\|\bm{v}_i - \bm{c}_i\|}
= \|\bm{v}_i - \bm{c}_i\| > 0.
\label{eq:projection_sign}
\end{equation}

Therefore, the sum becomes:
\begin{equation}
\sum_{j \in \mathcal{N}(i)} \bm{n} \cdot (\bm{v}_i - \bm{v}_j) = |\mathcal{N}(i)| \cdot \|\bm{v}_i - \bm{c}_i\| = |\mathcal{N}(i)| \cdot \|\bm{\kappa}_i\|.
\end{equation}

We now analyze the projection of the force $\bm{f}(i)$ onto the unit vector $\bm{n}$, given by:
\begin{equation}
\bm{n} \cdot \bm{f}(i) = \sum_{\bm{v}_j \neq \bm{v}_i} -\frac{CK^{p+1}}{\|\bm{v}_i - \bm{v}_j\|^p} \left( \bm{n} \cdot (\bm{v}_i - \bm{v}_j) \right)
+ \sum_{\bm{v}_j \leftrightarrow \bm{v}_i} \frac{\|\bm{v}_i - \bm{v}_j\|}{K} \left( \bm{n} \cdot (\bm{v}_i - \bm{v}_j) \right).
\label{eq:n_dot_f}
\end{equation}

To estimate the magnitude of this projection, we apply the reverse triangle inequality:
\begin{equation}
|\bm{n} \cdot \bm{f}(i)| \geq 
\left| \sum_{\bm{v}_j \leftrightarrow \bm{v}_i} \frac{\|\bm{v}_i - \bm{v}_j\|}{K} \left( \bm{n} \cdot (\bm{v}_i - \bm{v}_j) \right) \right|
- 
\left| \sum_{\bm{v}_j \neq \bm{v}_i} \frac{CK^{p+1}}{\|\bm{v}_i - \bm{v}_j\|^p} \left( \bm{n} \cdot (\bm{v}_i - \bm{v}_j) \right) \right|.
\end{equation}

Let $\lambda_{\min} = \min_{j \in \mathcal{N}(i)} \|\bm{v}_i - \bm{v}_j\| > 0$, where $\mathcal{N}(i)$ denotes the set of neighbors of $\bm{v}_i$. Since $\lambda_{\min}$ is the minimum distance among all neighbors, we have:
\begin{equation}
\|\bm{v}_i - \bm{v}_j\| \geq \lambda_{\min}, \quad \forall j \in \mathcal{N}(i).  
\end{equation}

This leads to the following lower bound:
\begin{equation}
\begin{aligned}
\sum_{\bm{v}_j \leftrightarrow \bm{v}_i} \frac{\|\bm{v}_i - \bm{v}_j\|}{K} \left( \bm{n} \cdot (\bm{v}_i - \bm{v}_j) \right)
&\geq \frac{\lambda_{\min}}{K} \sum_{j \in \mathcal{N}(i)} \bm{n} \cdot (\bm{v}_i - \bm{v}_j) \\
&= \frac{\lambda_{\min}}{K} |\mathcal{N}(i)| \cdot \|\bm{\kappa}_i\|.
\end{aligned}
\end{equation}

Thus, the absolute value of this term is bounded below by:
\begin{equation}
\left| \sum_{\bm{v}_j \leftrightarrow \bm{v}_i} \frac{\|\bm{v}_i - \bm{v}_j\|}{K} \left( \bm{n} \cdot (\bm{v}_i - \bm{v}_j) \right) \right|
\geq \frac{\lambda_{\min}}{K} |\mathcal{N}(i)| \cdot \|\bm{\kappa}_i\|.
\end{equation}

Now consider the second sum. We partition the summation over all $\bm{v}_j \neq \bm{v}_i$ into contributions from neighboring points ($j \in \mathcal{N}(i)$) and distant points ($j \notin \mathcal{N}(i)$):
\begin{equation}
\sum_{\bm{v}_j \neq \bm{v}_i} \frac{CK^{p+1}}{\|\bm{v}_i - \bm{v}_j\|^p} \left( \bm{n} \cdot (\bm{v}_i - \bm{v}_j) \right)
= CK^{p+1} \left(
\sum_{j \in \mathcal{N}(i)} \frac{|\bm{n} \cdot (\bm{v}_i - \bm{v}_j)|}{\|\bm{v}_i - \bm{v}_j\|^p}
+ \sum_{j \notin \mathcal{N}(i)} \frac{|\bm{n} \cdot (\bm{v}_i - \bm{v}_j)|}{\|\bm{v}_i - \bm{v}_j\|^p}
\right).
\end{equation}

Applying the triangle inequality gives the upper bound:
\begin{equation}
\left| \sum_{\bm{v}_j \neq \bm{v}_i} \frac{CK^{p+1}}{\|\bm{v}_i - \bm{v}_j\|^p} \left( \bm{n} \cdot (\bm{v}_i - \bm{v}_j) \right) \right| 
\leq
CK^{p+1} \left( 
\sum_{j \in \mathcal{N}(i)} \|\bm{v}_i - \bm{v}_j\|^{1 - p} 
+ 
\sum_{j \notin \mathcal{N}(i)} \|\bm{v}_i - \bm{v}_j\|^{1 - p} 
\right).
\end{equation}

For $j \in \mathcal{N}(i)$, we have $\|\bm{v}_i - \bm{v}_j\| \geq \lambda_{\min}$, so:
\begin{equation}
\sum_{j \in \mathcal{N}(i)} \|\bm{v}_i - \bm{v}_j\|^{1 - p} 
\leq |\mathcal{N}(i)| \cdot \lambda_{\min}^{1 - p}.
\end{equation}

For $j \notin \mathcal{N}(i)$ satisfying $\|\bm{v}_i - \bm{v}_j\| \geq \xi$, we get:
\begin{equation}
\sum_{j \notin \mathcal{N}(i)} \|\bm{v}_i - \bm{v}_j\|^{1 - p} 
\leq \frac{N}{\xi^{p - 1}}.
\end{equation}

Combining both bounds, we deduce:
\begin{equation}
\left| \sum_{\bm{v}_j \neq \bm{v}_i} \frac{CK^{p+1}}{\|\bm{v}_i - \bm{v}_j\|^p} \left( \bm{n} \cdot (\bm{v}_i - \bm{v}_j) \right) \right| 
\leq 
CK^{p+1} \left( |\mathcal{N}(i)| \cdot \lambda_{\min}^{1 - p} + \frac{N}{\xi^{p - 1}} \right).
\label{eq:beta_bound}
\end{equation}

In conclusion:
\begin{equation}
\|\bm{f}(i)\| \geq \|\bm{n} \cdot \bm{f}(i)\| \geq \frac{\lambda_{\min}}{K} |\mathcal{N}(i)| \cdot \|\bm{\kappa}_i\| - CK^{p+1} \left( |\mathcal{N}(i)| \cdot \lambda_{\min}^{1 - p} + \frac{N}{\xi^{p - 1}} \right).
\end{equation}

To simplify the expression, we define:
\begin{equation}
\alpha = \frac{\lambda_{\min} |\mathcal{N}(i)|}{K} > 0,
\end{equation}
\begin{equation}
\beta = CK^{p+1} \left( |\mathcal{N}(i)| \cdot \lambda_{\min}^{1 - p} + \frac{N}{\xi^{p - 1}} \right),
\end{equation}
where $\alpha$ and $\beta$ remain constant for a given point cloud.

Combining these with the previously established bounds, we obtain the final inequality:
\begin{equation}
\|\bm{f}(i)\| \geq \|\bm{n} \cdot \bm{f}(i)\| \geq \alpha \|\bm{\kappa}_i\| - \beta.
\end{equation}

\end{document}